\documentclass[%
 aip,
amssymb,amsthm,amsmath,
 reprint,%
]{revtex4-1}

\usepackage{graphicx}
\usepackage{dcolumn}
\usepackage{bm}
\usepackage{upgreek} 
\usepackage{hyperref}

\usepackage[utf8]{inputenc}
\usepackage[T1]{fontenc}
\usepackage{mathptmx}
\usepackage{xcolor}
\usepackage{xspace}
\usepackage{titlesec}

\titleformat{\paragraph}
{\normalfont\normalsize\bfseries}{\theparagraph}{1em}{}
\titlespacing*{\paragraph}
{0pt}{3.25ex plus 1ex minus .2ex}{1.5ex plus .2ex}

\newcommand{\umux}{$\upmu$mux\xspace}
\newcommand{\degr}{$^\circ$}
\newcommand{\Ohm}{$\Omega$\xspace}
\newcommand{\Order}{O}
\newcommand{\parthz}{pA/$\sqrt{\text{Hz}}$\xspace}

\begin{document}

\preprint{AIP/123-QED}

\title[SMuRF Tone-Tracking Electronics]{SLAC Microresonator RF (SMuRF) Electronics:\\A tone-tracking readout system for superconducting microwave resonator arrays}

\author{Cyndia Yu}
  \email{cyndiayu@stanford.edu}
  \affiliation{Department of Physics, Stanford University; Stanford, CA 94305; USA}
  \affiliation{Kavli Institute for Particle Astrophysics and Cosmology; Stanford, CA 94305; USA}
\author{Zeeshan Ahmed}%
  \affiliation{Kavli Institute for Particle Astrophysics and Cosmology; Stanford, CA 94305; USA}
  \affiliation{SLAC National Accelerator Laboratory; Menlo Park, CA 94025; USA}
\author{Josef C. Frisch}
  \affiliation{SLAC National Accelerator Laboratory; Menlo Park, CA 94025; USA}
\author{Shawn W. Henderson}
  \affiliation{Kavli Institute for Particle Astrophysics and Cosmology; Stanford, CA 94305; USA}
  \affiliation{SLAC National Accelerator Laboratory; Menlo Park, CA 94025; USA}%
\author{Max Silva-Feaver}
  \affiliation{Department of Physics, University of California San Diego; La Jolla, CA 92093; USA}
\author{Kam Arnold}
  \affiliation{Department of Physics, University of California San Diego; La Jolla, CA 92093; USA}
\author{David Brown}
  \affiliation{SLAC National Accelerator Laboratory; Menlo Park, CA 94025; USA}
\author{Jake Connors}
  \affiliation{National Institute of Standards and Technology; Boulder, CO 80305; USA}
\author{Ari J. Cukierman}
  \affiliation{Department of Physics, Stanford University; Stanford, CA 94305; USA}
  \affiliation{Kavli Institute for Particle Astrophysics and Cosmology; Stanford, CA 94305; USA}
\author{J. Mitch D'Ewart}
  \affiliation{SLAC National Accelerator Laboratory; Menlo Park, CA 94025; USA}
\author{Bradley J. Dober}
  \affiliation{National Institute of Standards and Technology; Boulder, CO 80305; USA}  
\author{John E. Dusatko}
  \affiliation{SLAC National Accelerator Laboratory; Menlo Park, CA 94025; USA}
\author{Gunther Haller}
  \affiliation{SLAC National Accelerator Laboratory; Menlo Park, CA 94025; USA}
\author{Ryan Herbst}
  \affiliation{SLAC National Accelerator Laboratory; Menlo Park, CA 94025; USA}
\author{Gene C. Hilton}
  \affiliation{National Institute of Standards and Technology; Boulder, CO 80305; USA}
\author{Johannes Hubmayr}
  \affiliation{National Institute of Standards and Technology; Boulder, CO 80305; USA}
\author{Kent D. Irwin}
  \affiliation{Department of Physics, Stanford University; Stanford, CA 94305; USA}
  \affiliation{SLAC National Accelerator Laboratory; Menlo Park, CA 94025; USA}
\author{Chao-Lin Kuo}
  \affiliation{Department of Physics, Stanford University; Stanford, CA 94305; USA}
  \affiliation{SLAC National Accelerator Laboratory; Menlo Park, CA 94025; USA}
\author{John A.B. Mates}
  \affiliation{National Institute of Standards and Technology; Boulder, CO 80305; USA}
\author{Larry Ruckman}
  \affiliation{SLAC National Accelerator Laboratory; Menlo Park, CA 94025; USA}
\author{Joel Ullom}
  \affiliation{National Institute of Standards and Technology; Boulder, CO 80305; USA}
\author{Leila Vale}
  \affiliation{National Institute of Standards and Technology; Boulder, CO 80305; USA}
\author{Daniel D. Van Winkle}
  \affiliation{SLAC National Accelerator Laboratory; Menlo Park, CA 94025; USA}
\author{Jesus Vasquez}
  \affiliation{SLAC National Accelerator Laboratory; Menlo Park, CA 94025; USA}
\author{Edward Young}
  \affiliation{Department of Physics, Stanford University; Stanford, CA 94305; USA}
  \affiliation{Kavli Institute for Particle Astrophysics and Cosmology; Stanford, CA 94305; USA}

\date{\today}

\begin{abstract}
We describe the newest generation of the SLAC Microresonator RF
(SMuRF) electronics, a warm digital control and readout system for
microwave-frequency resonator-based cryogenic detector and multiplexer
systems such as microwave SQUID multiplexers (\umux) or microwave 
kinetic inductance detectors (MKIDs).
Ultra-sensitive measurements in particle physics and astronomy 
increasingly rely on large arrays of cryogenic sensors, which in
turn necessitate highly multiplexed readout and accompanying
room-temperature electronics.
Microwave-frequency resonators are a popular tool for cryogenic
multiplexing, with the potential to multiplex thousands of detector
channels on one readout line.  
The SMuRF system provides the capability
for reading out up to 3328 channels across a 4-8 GHz bandwidth.
Notably, the SMuRF system is unique in its implementation of a
closed-loop tone-tracking algorithm that minimizes RF power transmitted
to the cold amplifier, substantially relaxing system linearity
requirements and effective noise from intermodulation products.
Here we present a description of the hardware, firmware, and software
systems of the SMuRF electronics, comparing achieved
performance with science-driven design requirements.
We focus in particular on the case of large-channel-count, low-bandwidth 
applications, but the system has been easily reconfigured for high-bandwidth 
applications.
The system described here has been successfully deployed in lab
settings and field sites around the world and is baselined for use on
upcoming large-scale observatories.
\end{abstract}

\maketitle

\section{\label{sec:intro}Introduction}
Superconducting detector technologies have enabled highly sensitive
measurements across a wide array of scientific applications, including
particle physics, astronomy, cosmology, materials science, chemistry,
biophysics, and quantum information
science\cite{enss05,irwinhilton,zmuidzinas12,sledgehammer,lee19,zadeh21,mariantoni11}.
As the instruments making these measurements move to ever-denser detector arrays to
increase their sensitivity or throughput, there is a growing need for advancements in
cryogenic multiplexing techniques to reduce thermal loading at
sub-Kelvin stages, cost, and integration complexity.
As an example, transition edge sensors (TESs) are a widely-used superconducting 
detector technology that have been fielded in arrays with over 10,000 
detector channels.\cite{spt3g} 
Cryogenic multiplexing of TESs on fielded experiments has been achieved with 
 time-division multiplexing (TDM)\cite{chervenak99}, frequency-division 
 multiplexing (FDM)\cite{dobbs04}, and Walsh code-division multiplexing
(CDM) techniques\cite{irwin10,morgan16}.
These technologies operate at $\sim$MHz frequencies and achieve
multiplexing factors of \Order(100) sensors read out per wire, but
are limited by high cost, complexity, and/or noise penalties as they move
towards higher multiplexing factors.
To achieve future science goals which benefit from larger and/or 
more tightly packed detector arrays, new strategies are being devised 
to push towards the much larger sensor counts required.

Microwave frequency-division multiplexing techniques offer one strategy 
for addressing these challenges by coupling signals to superconducting 
microresonators on a single RF feedline\cite{day03}.
The high quality factors achieved by standard superconducting
nanofabrication techniques along with the large readout bandwidth on
the microwave feedline in the 0.1-10~GHz frequency range enable
\Order(1000) multiplexing factors on a single coaxial line.
In microwave SQUID multiplexing (\umux), each detector is inductively
coupled via an RF SQUID to a unique GHz-frequency resonator such that
detector signals correspond to shifts in the resonator's frequency.
\cite{irwin04,mates08}
The resonator may also serve as both the readout and the detector
simultaneously, as is the case for microwave kinetic inductance
detectors (MKIDs), where incoming photons break Cooper pairs and
induce a modulation of the resonance transmission\cite{zmuidzinas12}.
In either scheme, many resonators with unique resonance frequencies 
can be fit in the available readout bandwidth, allowing large numbers 
of detectors to be read out on a single coaxial input/output pair.

Despite the promise of these microwave resonator based systems,
challenges in both cold and warm components had until recently limited 
these systems from achieving their full multiplexing potential.
Previous implementations of \umux have traditionally been limited by
resonator spacing and bandwidth achieved in fabrication.
These challenges have since been largely resolved, and resonators with
high quality factors, reproducible frequency spacing, and
bandwidths suitable for \Order(2000) multiplexing factors are now being 
consistently produced.\cite{dober21}
KID devices have similarly achieved sufficiently uniform fabrication 
and post-fabrication frequency adjustments to realize \Order(1000) channel 
multiplexing factor arrays.\cite{shu18,mckenney19}
For several microwave resonator-based readout systems such as bolometric 
applications of \umux, which operate 
at higher resonator drive powers, linearity of the cold amplifier and subsequent 
warm electronics have become important constraints in the achievable multiplexing 
factor. 

The SLAC Microresonator RF (SMuRF) warm electronics system has been
designed to meet these linearity challenges and read out up to 3328
channels in a 4~GHz band.
The SMuRF provides both the RF and low-frequency signals required
to enable full operation of a microresonator-based readout system.
Here and throughout the text, we refer to RF as the frequency regime 
spanning about 1-10~GHz, while DC refers to both DC and low-frequency 
(up to about 1~MHz) signals. 
The SMuRF system is unique in its implementation of fast tone tracking via a closed-loop 
adaptive filter, which thereby reduces power on the cryogenic
amplifiers and RF mixers.
This in turn increases the RF dynamic range of the electronics, 
allowing for improved RF linearity and increasing the achievable
multiplexing factor.
An earlier version of SMuRF was described demonstrating multiplexed 
readout of nearly 500 \umux channels.\cite{henderson18}
SMuRF has already been used to read out TES-coupled arrays in 
laboratory and field observing settings, achieving multiplexing factors of 
over 900 channels on a single coaxial pair.\cite{cukierman19,mccarrick21}
This system is being used for several upcoming Cosmic Microwave
Background (CMB) experiments including Simons
Observatory\cite{galitzki18} and BICEP Array\cite{moncelsi20}.

While SMuRF was designed and optimized for the read out of
TES-coupled \umux resonators, it is flexibly reconfigureable for
application to many other sensor and microwave readout technologies 
including KIDs, magnetic microcalorimeters (MMCs), and even potentially 
superconducting qubits~\cite{zmuidzinas12,stevenson19,chen12}.
Many other warm RF systems have been developed for GHz resonator read
out, particularly targetting the \umux and MKID applications.\cite{bourrion11,mchugh12,gordon16,vanrantwijk16,gard18,fruitwala20,smith21,stefanazzi22}
Compared to those systems, SMuRF offers significant advantages
including linear RF performance for high channel density over a 4~GHz bandwidth,
integrated high-performance low-frequency signal generation, timing and 
data streaming integration with large-scale experiment infrastructures, and a
unique tone-tracking capability.

In this paper, we present the design and performance of the current SMuRF 
electronics. 
We begin with a brief introduction to the superconducting resonator systems
that SMuRF has been designed to read out in
Section~\ref{sec:umuxintro}.
We pay special attention to \umux for large-format arrays to
motivate the design needs of the readout electronics.
In Section~\ref{sec:req} we outline the system requirements driven by
the science goals and design constraints in anticipated applications.
We follow with an overview of the system in Section~\ref{sec:over} and
descriptions of its hardware and firmware in Sections~\ref{sec:hard}
and~\ref{sec:firm} respectively.
We briefly describe the data streaming system, integration with
instrument control systems, and the end user software environment in
Section~\ref{sec:soft}.
Finally, we present the system performance in Section~\ref{sec:perf},
in particular demonstrating that the SMuRF system meets or exceeds its
design goals.

\section{\label{sec:umuxintro}Superconducting Resonator Readout}

There are many scientific applications that utilize
superconducting microwave resonators as the sensing and/or readout
element for ultra-sensitive circuits, including but not limited to
superconducting cryogenic sensor arrays for both low-bandwidth signals
and pulse detection, qubit control and readout, and quantum
information storage and
addressing.\cite{day03,wang09,hofheinz09,mariantoni11}
These applications have a large range of signal bandwidths ranging 
from <~1~Hz to 10s of kHz or faster. 
There is generically a tradeoff between signal bandwidth and channel 
count: to fit large numbers of sensors in the usable readout 
bandwidth the detectors must be bandwidth-limited, while larger-bandwidth 
detectors utilize a greater fraction of the total readout bandwidth and cannot 
be packed as tightly in frequency space. 
To maximize performance on the smaller number of channels being 
used in the higher-bandwidth cases, system linearity is more limited, meaning 
that the fewer channels may be run before performance begins to degrade. 

Since the RF linearity requirements for the high channel density case
drive the SMuRF hardware design in particular, we focus here on low
bandwidth, high-channel-count application, such as transition-edge
sensor (TES) bolometer or MKID arrays for astronomical sensing.
We further focus on the case of microwave SQUID multiplexers (\umux)
coupled to low-bandwidth TESs, since the higher resonator drive powers
and flux ramp demodulation present additional complications beyond KID
systems.
In this section we briefly review key terms pertaining to resonator
readout and give a brief overview of the TES-coupled \umux system and
typical integration to orient the reader for later sections.

\subsection{\label{sec:resintro}Microwave Resonators}

The microwave resonators used for cryogenic detector readout are 
coupled to a common RF transmission line to enable multiplexing.
The RF transmission line is excited with input microwave power 
to interrogate the state of each resonator simultaneously. 
This input power can be broadband noise, or as is more often the case for 
high-channel count systems with known resonance frequencies, probe tones 
tuned to each resonance frequency. 
We typically operate the RF line in transmission mode and adjust the power levels at the
resonator inputs with attenuation on the input and amplifiers on the output. 
In the discussion that follows we assume that the resonators are
coupled to the RF line such that they exhibit a dip in forward
transmission ($\mathrm{S_{21}}$) amplitude on resonance; other systems
are easily mapped onto this case.\cite{wang21}

\begin{figure}
    \includegraphics[width=0.99\columnwidth]{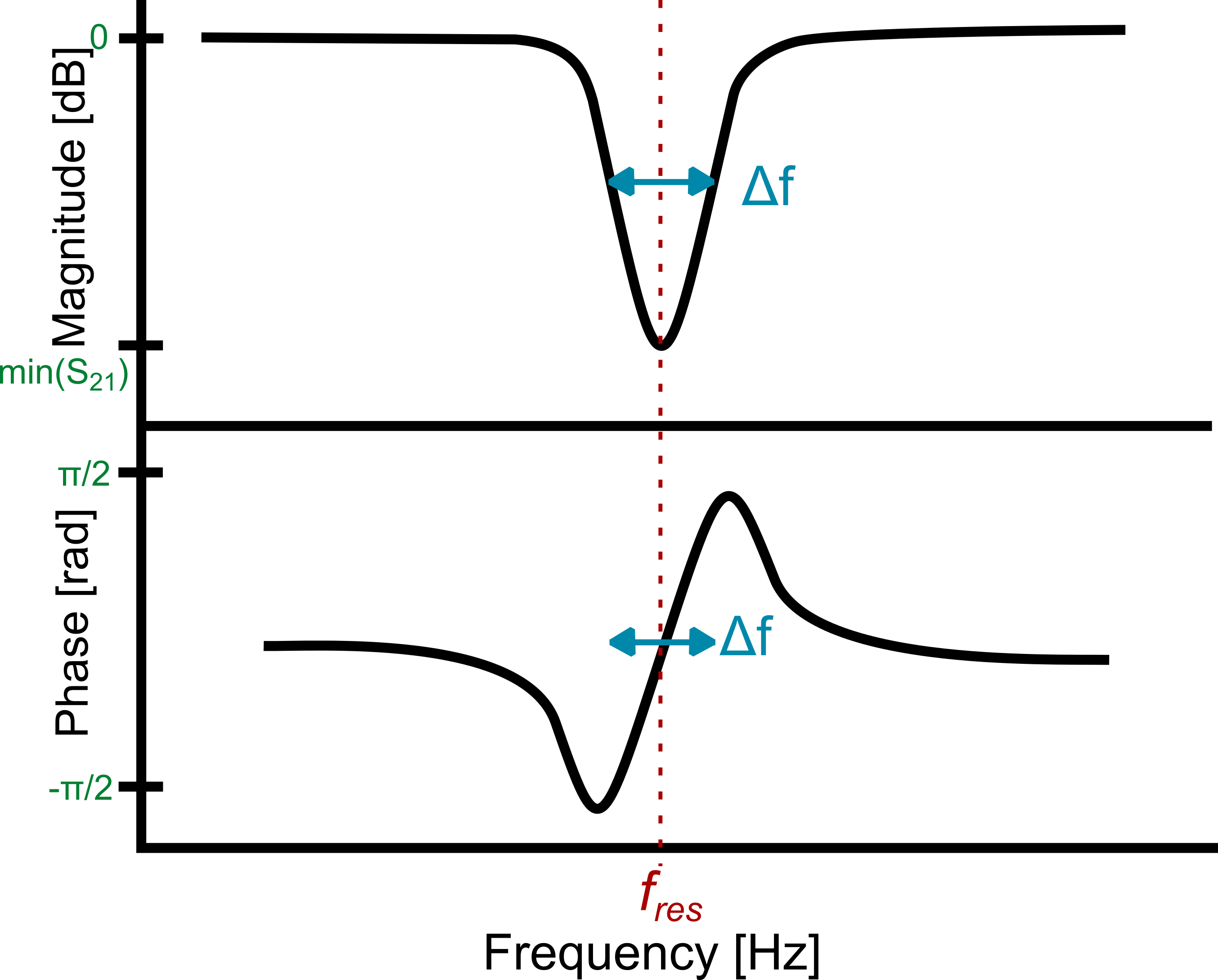}
    \caption{Example $\mathrm{S_{21}}$ amplitude [Top] and phase [Bottom] versus 
    frequency for an idealized resonator. The resonance frequency $f_\mathrm{res}$ 
    corresponds here to the minimum of transmission. The width characterizes the 
    sharpness of the resonance, which is inversely related to the quality factor $Q$. 
    }
    \label{fig:exs21}
\end{figure}

An example resonator $\mathrm{S_{21}}$ is plotted in Figure~\ref{fig:exs21}.
In this idealized case, the resonance frequency $f_\mathrm{res}$
corresponds to the minimum of the transmission amplitude at the center
of the resonance, though in general resonances can be asymmetric.
The complex phase undergoes a sign change through the resonance. 
The width of the resonance in frequency space is characterized by the
resonator quality factor $Q$, which is related to the width of the
resonance $\Delta f$ at full width half maximum by $Q = f_\mathrm{res}
/ \Delta f$.
We further differentiate between the internal quality factor denoted $Q_i$ where
$Q_i \equiv Q / \mathrm{S}_{21}^\mathrm{min}$ and the coupling quality 
factor $Q_c$, where the total $Q$ is related to the coupling and internal 
quality factors by $1/Q = 1/Q_c + 1/Q_i$.
The complex forward transmission through an idealized resonance may be 
parameterized as
\begin{equation}
\mathrm{S}_{21}(f) = 1 - \frac{Q}{Q_c}\frac{1}{1+2jQx}
\label{eq:s21}
\end{equation}
\noindent where $x = (f-f_\text{res})/f_\text{res}$ is the fractional detuning 
from the resonance frequency.\cite{zmuidzinas12}

\begin{figure}
    \includegraphics[width=0.99\columnwidth]{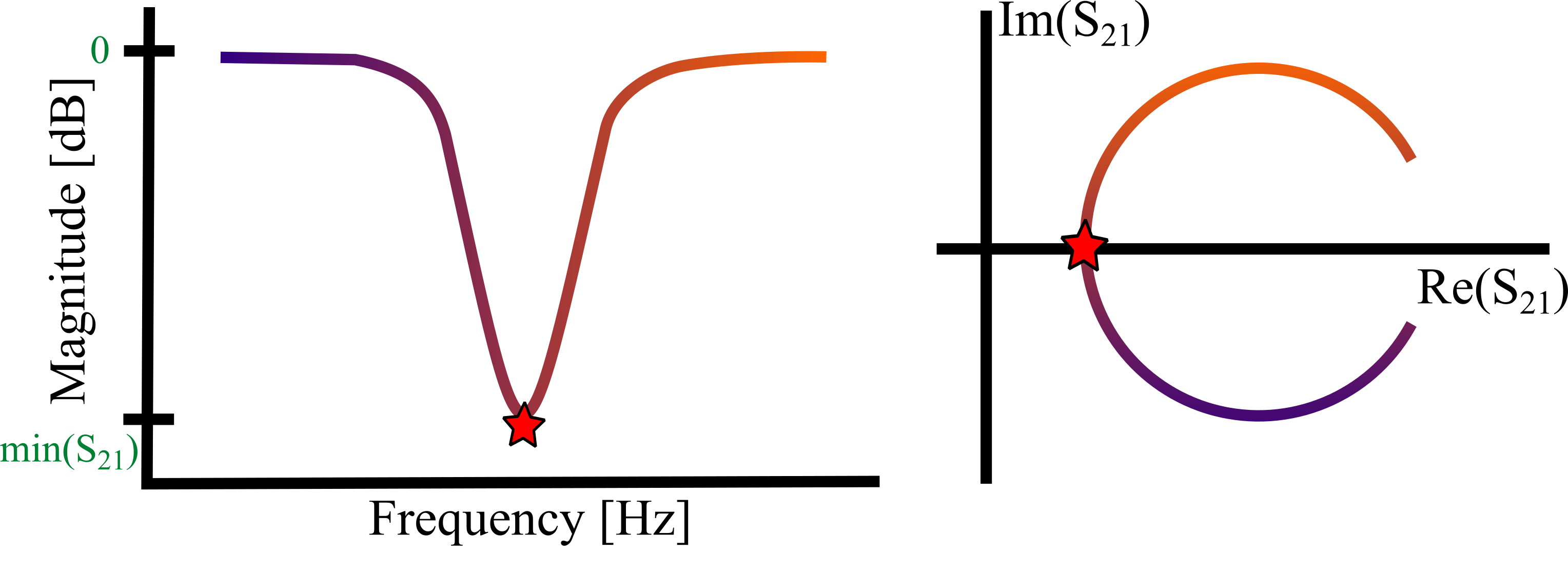}
	\caption{[Right] Example resonance circle corresponding to an ideal resonance, 
	with magnitude response [Left] for reference. The S$_{21}^\mathrm{min}$ is 
	denoted with a star in both cases. The color gradient denotes the same points in 
	frequency in both panels. We see that the resonance generically traces 
	out a circle in the complex plane, with the diameter of the circle corresponding 
	to the depth of the resonance. 
	}
    \label{fig:excirc}
\end{figure}

We can equivalently plot the S$_{21}$ from Figure~\ref{fig:exs21} in the 
complex plane, where the resonance traces out a circle with the resonance 
frequency at the point of intersection with the real axis as seen in 
Figure~\ref{fig:excirc}. 
For an ideal resonance, this circle is positioned such that on resonance the 
transmission is entirely real and small changes in the resonance frequency 
result in changes in transmission along the imaginary axis, or phase quadrature 
of the resonator. 
The real axis corresponds to the amplitude quadrature. 
The effects of cabling and loss, however, may in general displace, scale, 
and/or rotate the resonance circle to an arbitrary position in the complex plane. 
We may equivalently refer to the amplitude and phase quadratures of the resonator 
as $I$ and $Q$, respectively, which are related by the same rotation and scaling 
to the electronics I and Q quadratures.

\subsection{\label{sec:umuxspecintro}The Microwave SQUID Multiplexer}

\begin{figure}
    \includegraphics[width=0.99\columnwidth]{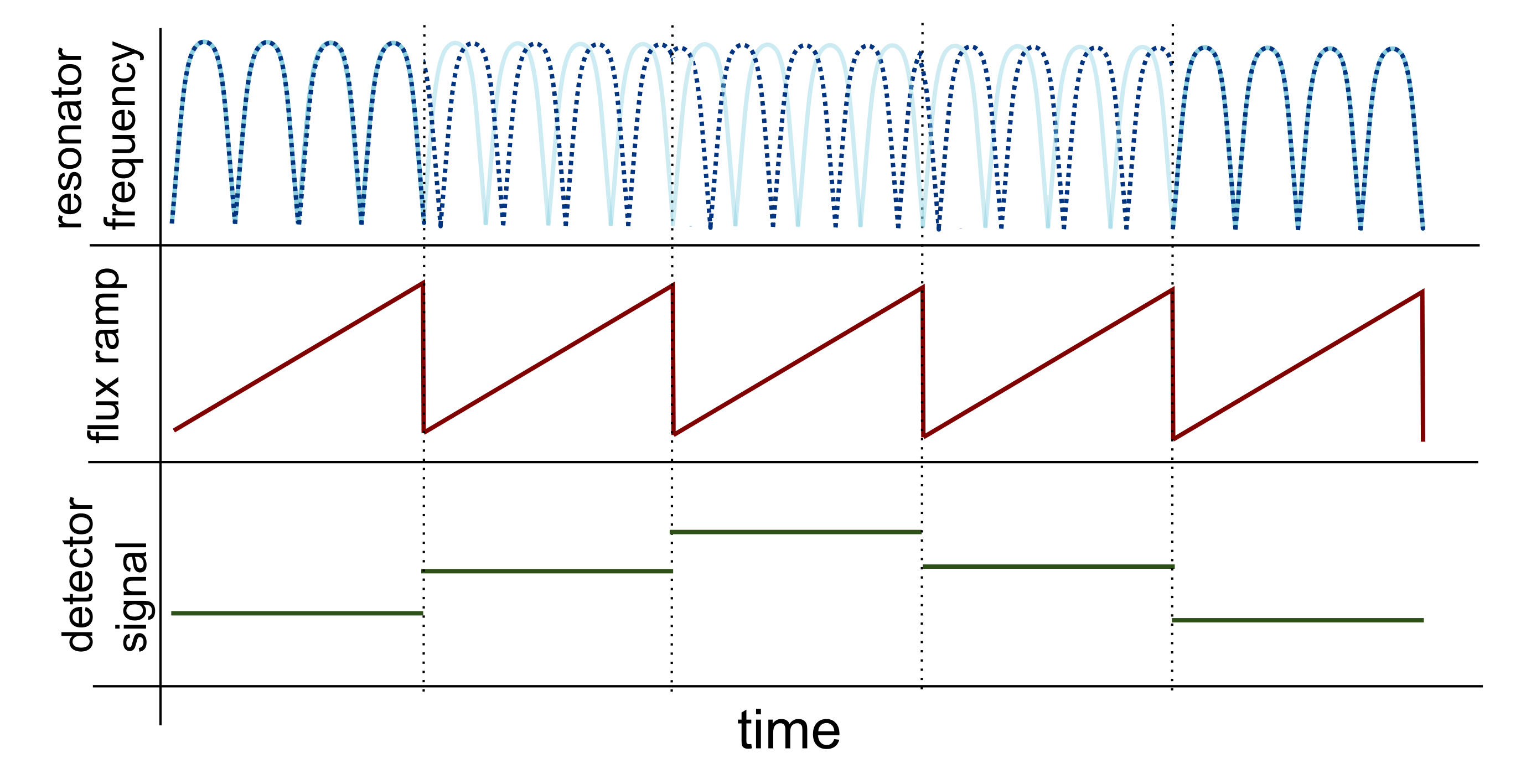}
    \caption{An overview of the phase modulation scheme of \umux. 
    The top panel depicts the resonance frequency, which shifts in response to 
    flux in the SQUID loop. The middle panel gives the flux ramp, which is typically 
    driven in a sawtooth pattern with amplitude sufficient to drive many flux quanta 
    through the SQUID (in this case, 4). The bottom panel gives an example detector 
    signal, which varies slowly compared with the flux ramp. The detector and flux ramp 
    signals both contribute flux to the SQUID; thus, the resonator response to the 
    detector signal is a shift in the phase of the flux ramp modulated frequency 
    shift (dark dotted lines on top panel) from the detector signal-free case 
    (light lines on top panel). }
    \label{fig:umuxphase_sketch}
\end{figure}

The microwave SQUID multiplexer (\umux) transduces a cryogenic detector signal, 
typically from a transition-edge sensor (TES) or a magnetic microcalorimeter (MMC), 
into flux in an RF SQUID loop.\cite{irwin04,mates08} 
Flux in the SQUID loop changes the effective inductance of the microwave resonator, 
causing a shift in its resonance frequency. 
A separate and common flux ramp bias simultaneously applied to all channels linearizes 
the SQUID response without the need for individual feedback lines.\cite{mates2012}
An incoming detector signal is therefore transduced into a phase shift in the SQUID 
response to flux ramp, which is periodic. 
A simplified picture of this phase modulation scheme is depicted in Figure~\ref{fig:umuxphase_sketch}. 

The SQUID flux to resonance frequency relation is a quasi-sinusoidal relation 
parameterized at low drive power by the peak to peak swing of the resonance 
frequency $\Delta f_\text{pp}$ and hysteresis parameter $\lambda$, where lower 
values of $\lambda$ correspond to more sinusoidal curves. 
It is commonly parameterized as\cite{mates2012}
\begin{equation}
\label{eq:freqvst}
\Delta f(t) = B\left(\frac{\lambda\cos(\omega_ct+\theta(t))}{1+\lambda\cos(\omega_ct+\theta(t))}\right)
\end{equation}
where $B$ is a constant related to $\Delta f_\text{pp}$. 
We note that Eq.~\ref{eq:freqvst} does not fully capture the complexity or 
dependencies of the SQUID curve, which is a subject of intense study.\cite{wegner22}

The SMuRF system hardware was designed for high linearity use cases with large channel counts. 
The current firmware was optimized for reading out microwave SQUID devices for CMB TES 
bolometer readout fabricated by the Quantum Sensing Group at NIST Boulder.\cite{dober21}
The design values for these resonators are resonator bandwidths of $\Delta f\sim 100~\mathrm{kHz}$, 
peak to peak frequency swing of $\Delta f_\text{pp}\sim 100~\mathrm{kHz}$, quality factor 
$Q_i > 100,000$, and $\lambda\sim 1/3$. 
They are optimally driven with input powers around $P_\mathrm{in}\sim -75\mathrm{dBm}$. 
The devices are being developed for the eventual goal of achieving a multiplexing factor 
of 2000 channels from 4-8~GHz on a single RF line. 
Other resonator technologies such as MKIDs, TKIDs, and even room-temperature resonators 
have been successfully read out with SMuRF systems as well, but did not drive design choices. 

The maximum multiplexing factor is set by the design value of the peak to peak swing 
$\Delta f_\text{pp}$ and the requirement that the resonances are spaced apart by a factor 
of 10 times $\Delta f_\text{pp}$ in frequency to avoid collisions and crosstalk. 
In general \umux devices have roughly matched resonance bandwidth and peak to peak 
frequency swings to maximize SQUID sensitivity without causing hysteresis or pathological 
SQUID responses.\cite{mates08}

Reconstruction of detector information is achieved by exciting the resonators with power 
and monitoring their state. 
The signal chain consists of a comb of microwave-frequency resonators capacitively coupled 
to a single RF transmission line. 
Each detector is uniquely identified by its resonance frequency through a 1:1 mapping 
between detectors and SQUID input coils. 
Since the initial resonance frequencies of the resonators can be measured in advance, 
the resonator excitation is often achieved with a comb of excitation probe tones, each 
tuned to a single resonator channel. 
In typical microwave resonator electronics systems, the resonator acts as a moving notch 
(narrow bandstop) filter that shifts off the probe tone, and each probe tone is monitored 
for changes in amplitude and phase. 
As will be discussed more in depth in \S~\ref{subsec:lin}, the RF linearity requirements 
on a system that must accommodate the possibility of several thousand off-resonance probe 
tones are stringent, motivating SMuRF's unique tone-tracking capability. 

\subsection{\label{sec:integration}Typical \umux System Implementation}
\begin{figure}
    \includegraphics[width=0.99\columnwidth]{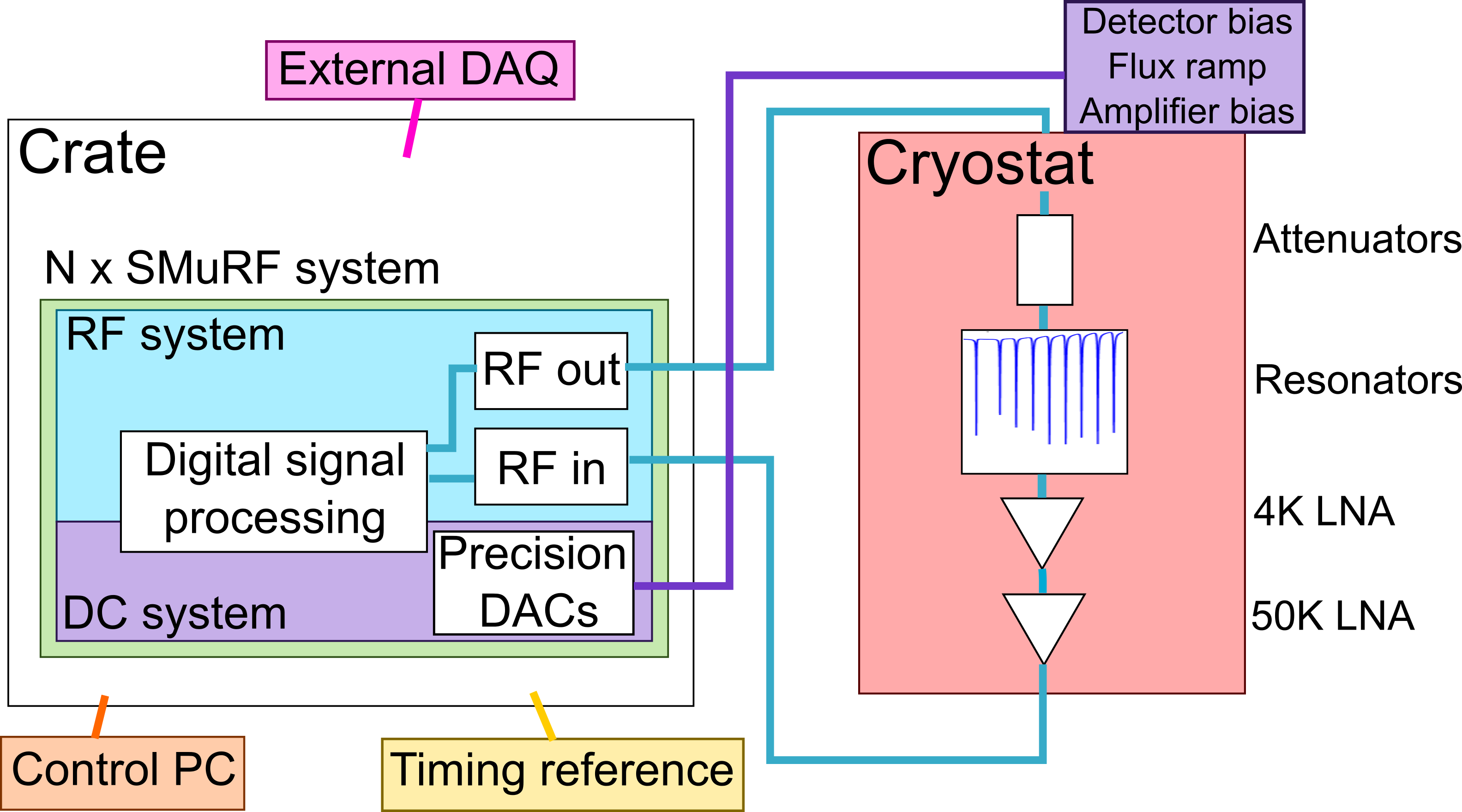}
    \caption{A typical integration between the SMuRF system and
      cryostat. The SMuRF system consists of RF and DC components that
      interact with the cryostat. The number of RF pairs may change
      depending on the multiplexing factor achieved by the cold
      devices. The DC lines supply the biases for RF amplifiers and
      detectors as well as the flux ramp.  The system additionally
      interacts with several peripherals, including an optional
      external timing reference and data acquisition system (DAQ). 
      Throughout this paper, the color scheme is kept broadly consistent 
      such that the main system computation and processing is given in 
      green, RF components are depicted in light blue, DC components in purple,
      data output in pink, timing in yellow, the control in orange,
      and the cryostat and other externally derived parameters in
      red.
    }
    \label{fig:integration}
\end{figure}

In this subsection we describe a typical implementation of a \umux system 
to motivate the electronics and peripheral system designs. 
The superconducting detector arrays are mounted at the cold stages of a 
cryostat and connected to room-temperature electronics systems via cabling 
that is passed through the cryostat vacuum feedthrough. 
A sketch of a typical system integration is given in Fig.~\ref{fig:integration}. 

The SMuRF electronics interface with a cryostat through both RF and DC connections. 
The RF connections consist of two RF ports per transmission line, one each for input and output. 
The DC connection is made with a single multipin hermetic connector, which is carried to 
colder temperature stages by a wire loom that is broken out at the 50K and 4K 
temperature stages to provide amplifier biases and filtering of the 
flux ramp and detector bias lines. 

The RF lines provide the signals that are carried via coaxial cables to the resonator 
transmission line. 
After interacting with the resonators, the transmitted power is amplified with one or 
more cryogenic RF amplifiers before being returned to the warm electronics for processing. 
In \umux, the resonators are coupled to the TES detectors, which must be biased for 
operation. 
In addition to the TES bias lines and RF amplifier bias lines, the DC lines 
must also supply a flux ramp line to the SQUIDs.

\section{\label{sec:req}Requirements}

The SMuRF system and auxiliary components must satisfy many performance criteria in 
order to meet the design specifictions of the experiments they are coupled to. 
Here, we outline briefly the system requirements driven by the expected functional 
requirements in \S~\ref{sec:reqcon}, the science goals in \S~\ref{sec:reqsci}, and 
the user needs in \S~\ref{sec:reqdes}. 
These design requirements are ultimately driven by the goal of reading out 2000 
\umux-coupled TES channels in a 4-8~GHz bandwidth, each with $\sim$-75~dBm input 
power, while maintaining total readout noise subdominant to the detector noise for 
a small-aperture telescope with no polarization modulation. 

\subsection{\label{sec:reqcon}Functional Requirements}
We first list the signals and connections that the SMuRF system is expected to provide,
based on the typical system implementation discussed in \S~\ref{sec:integration}.

The RF system of SMuRF must provide one RF signal generation and readback pair per 
cryostat coaxial line. 
The multiplexing factors achieved by \umux are set by the cryogenic devices, which 
are currently being designed for channel densities of about 500 resonators per 1~GHz 
of bandwidth across the 4-8~GHz range. 
For 1000x multiplexing factors we thus require two pairs of coaxial cables, while for 
2000x multiplexing factors only a single pair is needed. 
Based on NIST CMB-style \umux design specifications, we require that SMuRF can generate 
up to 2000 probe tones at about $\sim$-35~dBm each across 4~GHz of total bandwidth, allowing 
for cold attenuation prior to the resonator input. 

For science operations, the TESs are usually operated with a constant
low-noise voltage bias. 
In calibration mode and to drive the detectors out of their superconducting 
state, a higher bias level must be supplied, though this state is subject to 
looser requirements on the noise performance. 
Depending on the design and operating conditions of the TES, the bias line 
may be required to supply between $\sim$10~$\mu$A and $\sim$5~mA 
of current.

The electronics should additionally provide low-noise DC biases for the cryogenic 
amplifiers. 
Previous measurements indicate that cascaded cryogenic amplifiers offer improved 
IP3 for minimal noise temperature penalty compared with typical noise in the 
cryogenic resonators.\cite{henderson18} 
In this case, a lower gain, low noise 4K amplifier operates as the first-stage 
amplifier followed by a 50K amplifier, which may operate at higher powers and 
thus offers better linearity than a single 4K amplifier alone.  
For other applications, a single LNA may be sufficient. 
The LNAs are biased with a gate and optional drain DC line that supply no more 
than $\pm 2$~V and \Order(10) mA. 
We thus require the electronics supply bias lines for up to two cryogenic amplifiers 
per transmission line, which we refer to throughout the text as located at 4K and 
50K, though they are not required to be fixed at these temperature stages. 

Beyond these DC signals, the electronics must include a low-frequency signal 
generation system that supplies the flux ramp line common to all resonators on 
the same coaxial line. 
To modulate the TES signal above the $1/f$ knee of superconducting resonators\cite{zmuidzinas12}, 
we require that the flux ramp line achieve at least 10~kHz $\Phi_0$ rate, 
defined as the sawtooth reset rate multiplied by the number of flux quanta swept per ramp. 
The required amplitude of the flux ramp depends on the mutual inductance to 
the SQUID, but must be able to sweep several $\Phi_0$. 
Assuming there is no current division in the cryostat to suppress noise, this 
places the minimum requisite current at about $\sim$10~$\mu$A for current device 
designs, which have couplings between the flux ramp coil and SQUID of about 
\Order(30)~pH.\cite{mates2012,dober21}.

To mitigate against unwanted pickup from ambient radio frequency interference 
(RFI), cables carrying critical low frequency bias signals between the warm 
electronics and cryostat input connectors must be well-filtered and kept as 
short as feasible. 

The electronics should interface with a control computer, allowing the user to 
command the system and receive data. 
The system should further provide the ability to interface with external timing systems 
such as provided from experiment or observatory control. 
While standalone laboratory test systems may be able to stream data directly to disk, 
we require that the SMuRF system be capable of interfacing with external data 
acquisition systems as required by large-scale experiments. 
We similarly require that a SMuRF system be capable of communicating with and operating 
alongside other SMuRF systems.

\subsection{\label{sec:reqsci}Science-driven Requirements}

\subsubsection{\label{subsec:noise}Noise} 

The ultimate figure of merit for the readout noise performance is the
noise referred to the detector input.
The noise performance has two components: the white noise and the
low-frequency behavior.
Achieving good noise performance places stringent constraints on all
of SMuRF's subsystems, including but not limited to: RF tone
generation and channelization, detector bias drives, flux ramp, and
all digital signal processing algorithms.
Since low-frequency noise on the detector bias line directly contributes 
low-frequency noise on the detectors, noise performance and stability of the
low-frequency subsystems is particularly crucial. 
\\

\paragraph*{White Noise}

\noindent The white noise requirement of the readout, which includes both 
the cryogenic readout devices and warm electronics, is set by the
requirement that it be subdominant to detector noise referred to the input.
For uncorrelated noise contributions to the total detector noise the 
various noise components add in quadrature;
thus, in order to have minimal impact we require the readout
electronics induce no more than 10\% penalty to the detector noise. 
CMB TES detectors are currently photon-noise limited, with noise-equivalent 
powers of \Order(40) aW/$\sqrt{\mathrm{Hz}}$ at 95~GHz.\cite{b3instrument}
We operate in the regime that total readout noise is subdominant to phonon and 
Johnson noise of the TES, which are in turn subdominant to photon noise. 
The amplitude of the TES photon noise-equivalent current is set by the operating
resistance, which varies widely across experimental configurations
from about 1~m\Ohm to 30~m\Ohm for DC-biased systems.\cite{mccarrick21,bk_tes}
For CMB detectors with antenna responses centered at 95~GHz, this 
translates roughly to \Order(100-160)~\parthz for a BICEP-like experiment 
with large operating resistance and Simons Observatory-like 
experiment with a low operating resistance, respectively.\cite{b3instrument,mccarrick21}

To maintain low total readout noise, we require that the added white noise 
due to the warm electronics be subdominant to the intrinsic noise of the 
superconducting resonators and cryogenic amplifiers, whose combined noise 
in turn is required to be subdominant to the detected signals. 
For NIST CMB-style \umux resonators, the dominant readout noise sources include 
two-level systems (TLS) in the resonator dielectric, HEMT noise, and DAC 
noise.\cite{mates11}
To incur no more than 10\% penalty in the total current noise, this places 
an overall requirement on the readout at roughly \Order(45-70)~\parthz, 
which includes both the cryogenic devices and warm electronics.

For the purposes of setting a requirement on the warm electronics, we must 
relate the detector input-referred current noise to more practical units. 
Here we define several quantities relating to RF electronics systems 
that set the engineering specifications for the SMuRF systems. 

\begin{figure}
\includegraphics[width=0.95\columnwidth]{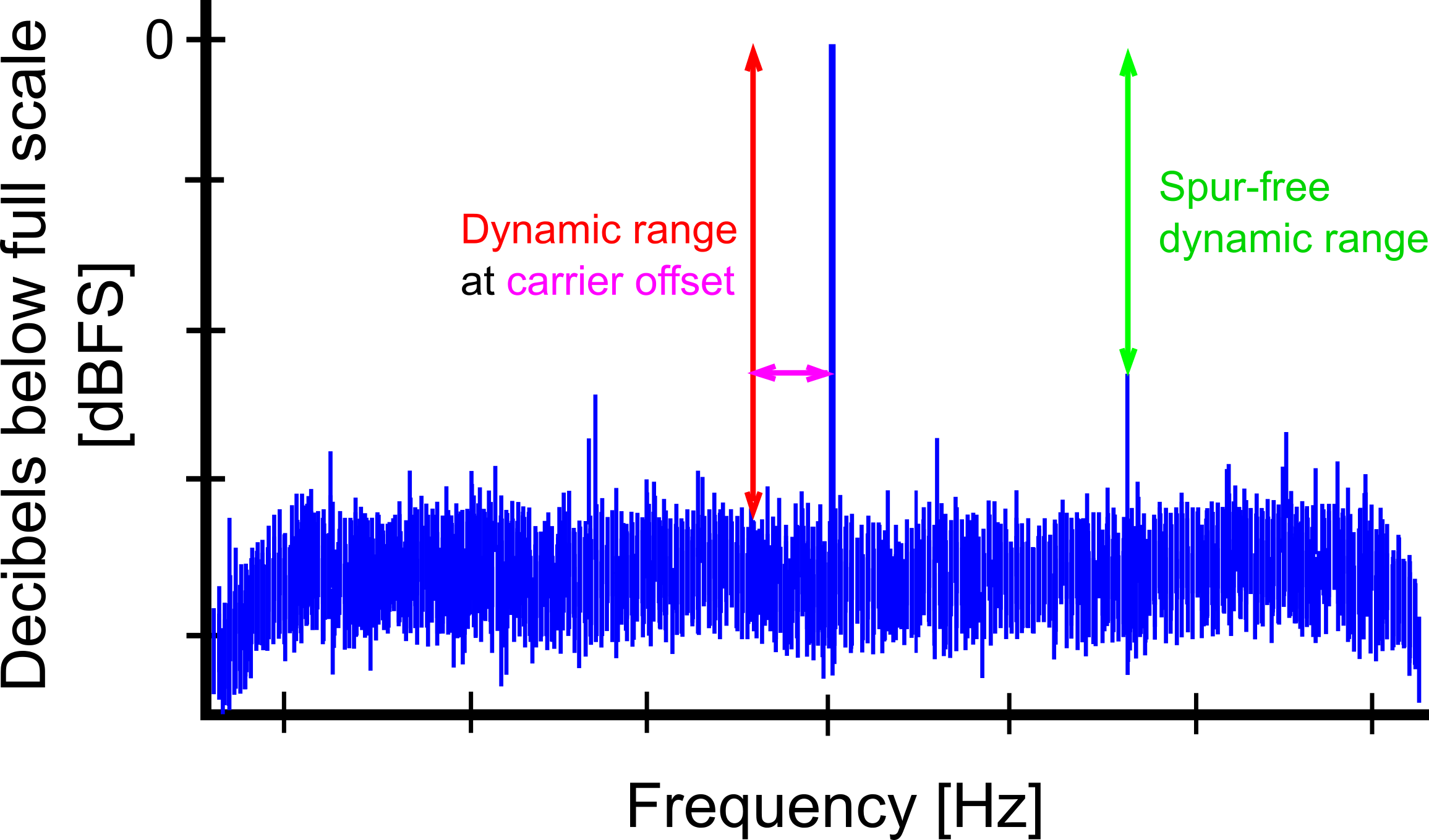}
\caption{A sketch of a carrier tone roughly centered within the displayed 
bandwidth to illustrate various terms relating to dynamic range. 
The dynamic range is measured at some offset frequency from the main carrier 
and represents the strength of the carrier signal relative to the noise 
floor. 
The noise floor may contain narrow-band spurs; thus, the spur-free dynamic 
range represents the strength of the carrier relative to the largest 
spur within the bandwidth. 
}
\label{fig:dynrange_sketch}
\end{figure}

In \umux systems, detector signals are modulated into the sidebands
of the drive tones at an \emph{offset frequency} given by the product of the
flux ramp sawtooth rate and the number of RF SQUID $\Phi_0$ swept in each
ramp, typically 4-6~kHz and 4-6 $\Phi_0$ respectively to get above the strong 
$1/f$-like component of the resonator noise. 
The difference between the power of the drive tone and the noise floor
generated in the sidebands of the tone by the warm electronics at this offset
frequency determines the maximum usable RF \emph{dynamic range}. 
The dynamic range is often expressed in units of dBc/Hz, or decibels relative 
to the carrier integrated over the measurement bandwidth, where the carrier is 
taken as the resonator probe tone and the measurement bandwidth is typically 
1~Hz. 
Alternatively, we can assign the maximum power of the carrier or ADC as ``full 
scale'' and express power levels relative to this maximum in terms of decibels
relative to full scale (dBFS). 
While we are typically interested in the dynamic range above the effective 
noise floor, we may at times consider the strength of the carrier signal above 
the strongest spurs, or the \emph{spur-free dynamic range} (SFDR). 
A sketch of these concepts for an arbitrary carrier and noise floor is given in 
Figure~\ref{fig:dynrange_sketch}. 

The maximum achievable dynamic range is limited by both noise and
nonlinearity in the active devices used to amplify, up- and down-mix,
and read out the tones.
In general there are tradeoffs between noise and RF nonlinearity; while
lowering drive power may improve the noise due to intermodulation
products produced by device nonlinearities, the dynamic range is eventually 
limited by the thermal noise floor of the electronics. 
Conversely, improving signal to noise over the thermal noise floor
with higher drive powers eventually results in an increased effective noise floor
due to intermodulation products of the probe tones.
We operate in the regime that the electronics dBc/Hz is not changed by
attenuation, as is generally the case for broadband RF electronics in
this parameter space.

The referral from electronics noise in dBc/Hz to TES noise equivalent
current involves converting noise in electronics phase to resonator frequency
noise via resonator and SQUID device parameters and referring to 
detector input-referred noise via factors of SQUID couplings. 
The full derivation and measurement is deferred to a separate publication. 
To remain subdominant to the intrinsic noise of NIST CMB-style 
\umux resonators, the requirement for the SMuRF electronics RF dynamic range per tone 
is -100~dBc/Hz at 30~kHz offset frequency when operating 2000 simultaneous channels.
This channel density goal is driven by the maximum expected density of 
the cryogenic multiplexer. 

The input-referred noise in the TES detector bias line is current divided between the 
shunt resistor and the TES. 
The ratio of the shunt resistance to the TES resistance under typical 
operating conditions is about~0.1. 
This sets the requirement on the noise on the bias line to 10 times the requirement 
on the current noise of the readout; for example, if the detector input-referred 
electronics noise requirement is 10~\parthz, then the noise on the 
TES bias line must be $\lesssim 100$~\parthz. 
The current noise requirement for a given application can vary widely 
depending on the TES operating resistance; but we require that the bias 
line current noise be $< 50$~\parthz, which is conservative for most 
DC voltage-biased TES arrays. 

The flux ramp has a smaller mutual inductance on the SQUID input coil 
than the TES, typically by a factor of 10. 
In order to further reduce noise, it may be current divided inside the 
cryostat. 
We assume no current division in the worst case and the same conservative 
5~\parthz total electronics noise requirement, resulting in a flux ramp 
current noise requirement of $< 50$~\parthz. 

\paragraph*{Low-Frequency Noise}

\noindent For mm-wave ground-based bolometric applications in particular, 
minimizing the $1/f$ noise of the detector time-ordered data is critical.\cite{bk3}
The conversion between angular scale on the sky and time-ordered data frequency 
depends on the telescope scan strategy. 
CMB ground-based surveys targeting large-scale sky features, such as the 
BICEP/Keck program, require time-ordered data that is white down to 
$\sim$~0.1~Hz. 
Thus, in the absence of additional polarization modulation we target the 
$1/f$ knee of the electronics to be  $\sim$~0.01~Hz. 
This low-frequency performance encompasses both the RF and DC subsystems, 
the latter presenting a particularly challenging constraint since any low 
frequency noise on the detector bias line directly translates to additional 
low frequency noise on the detector. 

The primary components of the SMuRF system which drive the low-frequency noise 
requirement are, in approximate order of impact, (1) the detector bias circuit, 
(2) the flux ramp circuit, (3) the RF signal circuit, and (4) the RF amplifier circuit.
In particular, the detector bias circuit is critical since fractional
drifts in detector bias current couple directly into fractional drifts
in the mean detector signal and gain in the high loop gain limit of
TES responsivity.

The low frequency drifts in the SMuRF-provided bias current are
sourced by intrinsic $1/f$ noise of the DACs and op-amps used in the
bias circuit and by the coupling between bias circuit
components (current-setting resistors and DAC voltage references)
and the temperature of their environment.
We choose DACs and op-amps such that their intrinsic $1/f$
contribution is subdominant to the temperature coupling effects.
In this limit the dominant driver of $1/f$ noise is the temperature
dependence of the passive circuit components.
These components must maintain an $f_\text{knee}\sim
0.01$~Hz even when coupling to the extreme case of diurnal ambient
temperature variations in the Atacama Desert where many CMB
observatories are currently operating or being built, assuming no active 
temperature control.
This sets the requirement for the combined temperature coefficient of all
components in the TES bias circuit to be 
$\lesssim 20~\mathrm{ppm}/^{\circ}\mathrm{C}$.

The flux ramp is typically operated at 10s of kHz, allowing for some relaxation 
of $1/f$ requirements on individual components. 
Since the flux ramp demodulation scheme relies on knowledge of the average flux 
ramp rate over long periods, the stability of this circuit is crucial. 
We find that the required component temperature stability and noise performance 
are comparable or  subdominant to that of the detector bias circuit; thus, 
choosing similarly rated components for both circuits suffices to meet this 
requirement.

The RF signal circuit low-frequency noise performance, defined here as the 
dynamic range at small carrier offset, is set by the stability of 
the local oscillator (LO), discussed further in \S~\ref{sec:amcrf}.
Due to the flux ramp scheme of \umux, the requirements on low-frequency 
noise in the  RF circuit and low-frequency variation of the LO position 
itself are not as stringent, and are subdominant to other sources of $1/f$ noise.

The SMuRF provides circuitry for cryogenic RF amplifier biasing, as
outlined in \S~\ref{sec:cryocard}.  These amplifiers are typically
sourced commercially and are robust to gain fluctuations
due to noise in the bias circuit.
Flux ramp modulation is expected to further suppress the majority of $1/f$ noise.
The phase modulation scheme of \umux additionally offers some amount
of immunity to overall amplitude fluctuations resulting from amplifier
gain noise.

\subsubsection{\label{subsec:lin}RF Linearity} 
Here, we set requirements on the system to minimize 3rd-order intermodulation 
products that degrade the RF dynamic range. 
The full octave of readout bandwidth used for 
this application places stringent requirements on the bandwidth of 
the RF components in the amplification and signal processing chain. 

For \emph{fixed-tone readout} where the RF probe tone is fixed 
while the resonance is modulating, thousands of unattenuated tones
are incident on the cryogenic amplifier as the resonances move in frequency. 
This is a particular challenge for \umux systems both due to the higher 
probe tone powers and since the resonance swings by \Order(1) resonator 
bandwidth in operation. 
To accommodate this large change in power, the cryogenic amplifier must 
trade off noise temperature, thus degrading the overall readout noise. 
The interaction of the RF probe tones further 
generates third-order intermodulation products in-band;
for 2000 RF tones in one octave of bandwidth there are over one billion third
order intermodulation products whose total power grows as the cube of
the fundamental tone power.
These intermodulation products combine and form an irreducible pseudo-noise
floor that degrades the dynamic range of the probe tones and 
may exceed the -100~dBc/Hz requirement discussed above,
adding to the detector-referred readout noise. 
In addition to sourcing noise, intermodulation products induce
crosstalk between channels that poses potential systematics for
science analysis.\cite{mates19}

SMuRF addresses this linearity problem with a \emph{tone-tracking algorithm},
which follows the modulating resonance with a closed-loop adaptive 
filter-based feedback.
Previous results demonstrate that this tone tracking is sufficient to
add negligible noise penalty when running \Order(400)
channels simultaneously versus a single channel at a time for
NIST CMB-style \umux resonators.\cite{henderson18,dober21}
The impact of adding additional channels is dependent on resonator
properties, most crucially the resonator dip depths.
For the full system we design with the goal of sufficient linearity
such that there is minimal penalty incurred by additional channels for
up to 2000 NIST CMB-style \umux channels on an RF line.
This sets a stringent requirement in particular on the RF components
including the ADCs, DACs, and RF mixers, as discussed in
\S~\ref{sec:amcrf}.

We emphasize that there is generically a tradeoff in RF systems between
noise temperature and linearity; the noise may be improved at the
expense of being able to operate fewer channels simultaneously.
For the SMuRF system the large channel counts
necessitated aggressive optimization for linearity at scale, but the noise
performance of the SMuRF system for a small number of channels can be
improved.

\subsubsection{\label{subsec:bw}Bandwidth} 

The required sampling bandwidth for the readout electronics is driven by the bandwidth 
of the detectors and the bandwidth of the resonator.
For CMB measurements, the science readout bandwidth is typically in the 0.1-300~Hz
range.\cite{dobbs04,henderson16}
This gives us a minimum bandwidth for reading
out TES bolometers of \Order(1~kHz) per
detector channel across a minimum of 2000 channels.
The detector noise itself extends higher in frequency; to 
keep the TESs stable they are designed to have bandwidth in the $\sim$~kHz 
range.\cite{irwinhilton}

For applications that require the reconstruction of fast pulses such 
as calorimetry, the energy resolution is set by the rise time and
height of the pulse.\cite{fowler16}
In this case, the SMuRF must be reconfigured for lower channel count,
higher bandwidth channelization and signal processing.
Further discussion of SMuRF optimization for pulse detection and
processing is deferred to a future publication.

The requisite TES readout bandwidths stated above are for the flux ramp-demodulated 
detector data, whose sample rate is set by the flux ramp rate. 
While the tone-tracking ability of SMuRF allows for readout of resonator modulation 
faster than the resonator bandwidth, it is still useful to use the resonator bandwidth 
as a heuristic for required per-channel readout bandwidth. 
The resonator bandwidth for NIST CMB-style \umux resonators is designed to be 
$\sim 100~\mathrm{kHz}$, well above the TES bandwidth. 
To avoid aliasing penalties and get above the $1/f$ noise from two-level systems 
(TLS) in the resonator, we flux ramp-modulate the signal to \Order(10~kHz). 
For the SMuRF to be capable of resolving \Order(10) kHz flux ramp rates, the 
per-channel digitization bandwidth requirement is set at \Order(1 MHz) to ensure 
enough samples per flux ramp cycle to allow for demodulation. 

Additional bandwidth and aliasing considerations of the \umux devices are discussed 
in a separate publication.\cite{ltd21}
For the purposes of checking performance for CMB-style resonators, we require 
that the system is capable of resolving and demodulating waveforms with frequencies 
up to $\sim 1~\mathrm{kHz}$. 

\subsubsection{\label{subsec:ct}Crosstalk} 
Crosstalk between channels contributes to unwanted signal that biases scientific 
analyses.
For CMB polarization analysis, the crosstalk between channels is held below 0.3\%.\cite{bk3}
The sources of signal crosstalk in the \umux cryogenic subsystem have been investigated 
and largely mitigated through cryogenic device design.\cite{mates19}
Most pertinently for SMuRF, the noise floor induced in part by the higher order 
intermodulation products of tones contributes some amount to an ``all-into-all'' 
form of crosstalk. 
We expect that the RF linearity requirement outlined in \S~\ref{subsec:lin} is 
sufficient to address this form of crosstalk. 
In particular, the -100~dBc/Hz for 2000 tones requirement translates to a maximum 
-43~dBc leakage between arbitrary channels in 1~GHz bandwidth. 
To be conservative, this sets a minimum requirement of $-60$~dB isolation between 
channels in any given 1~GHz bandwidth.

Beyond the linearity requirement, there may remain some amount of
crosstalk between channels in the electronics.
The NIST CMB-style \umux resonator is expected to dominate any crosstalk 
contribution from the SMuRF electronics.\cite{dober21}
We thus require simply that the electronics-induced crosstalk be subdominant
to the crosstalk induced by the cryogenic resonators.

\subsection{\label{sec:reqdes}Ancillary Design Requirements}
We note here several additional design constraints guided by the unique operating 
environments encountered by SMuRF and similar microwave resonator-based systems. 
\begin{itemize}
\item\textbf{Environment: }
The SMuRF electronics are expected to operate in a wide range of ambient temperature 
conditions without noise penalty. 
In the worst case, the electronics are air-cooled in ambient conditions. 
Additionally, many CMB experiments operate at high altitudes where oxygen levels 
are substantially lower than sea level. 
Extra care must be taken to ensure that the electronics do not overheat or otherwise 
fail in these conditions. 
The SMuRF was designed to operate at altitudes up to 5200 meters above sea level with 
ambient temperatures below 25\degr C. 
\item\textbf{Power: }Each SMuRF system's maximum power consumption
  must fit within the power budget of the intended application.  For
  high altitude astronomical observatories which often operate on a
  spare total power budget, this poses a significant constraint due to
  the high signal and switching speeds, the computationally intensive 
  nature of the SMuRF processing algorithm, and the power demands of the 
  electronics crate and peripherals. For reference, the total electrical 
  power available for readout for current sites is about \Order(1-3)~kW 
  for 20,000 channels. 
\item\textbf{Vibration Sensitivity: }The detector readout electronics on telescopes 
is often mounted as comoving with the receiver, which for CMB applications scans the 
sky at several degrees per second and turnaround accelerations of \Order(1)\degr/s$^2$. 
The SMuRF should be able withstand this movement without damage or noise penalty. 
In addition, multiple SMuRF systems are often run together in adjacent slots of one 
electronics crate. 
The presence of additional SMuRF systems in adjacent slots should not impact the 
performance of any given system. 
\item\textbf{Modularity: }Components of the system should be designed to allow for 
separate testing and integration, allowing for swapping and upgrading parts as necessary 
on deployed systems. 
\item\textbf{Ease of Use: }The user-facing interface and software should be accessible 
to end users who may desire to use SMuRF as a readout platform without expert knowledge 
of the electronics, firmware, or cryogenic multiplexer. This drives the SMuRF design 
towards modular components that are easily swappable, open source software, and 
compatibility with standard scientific laboratory computers and operating systems. 
\end{itemize}

\section{\label{sec:over}SMuRF Electronics Overview}
\begin{figure*}
    \includegraphics[width=0.9\textwidth]{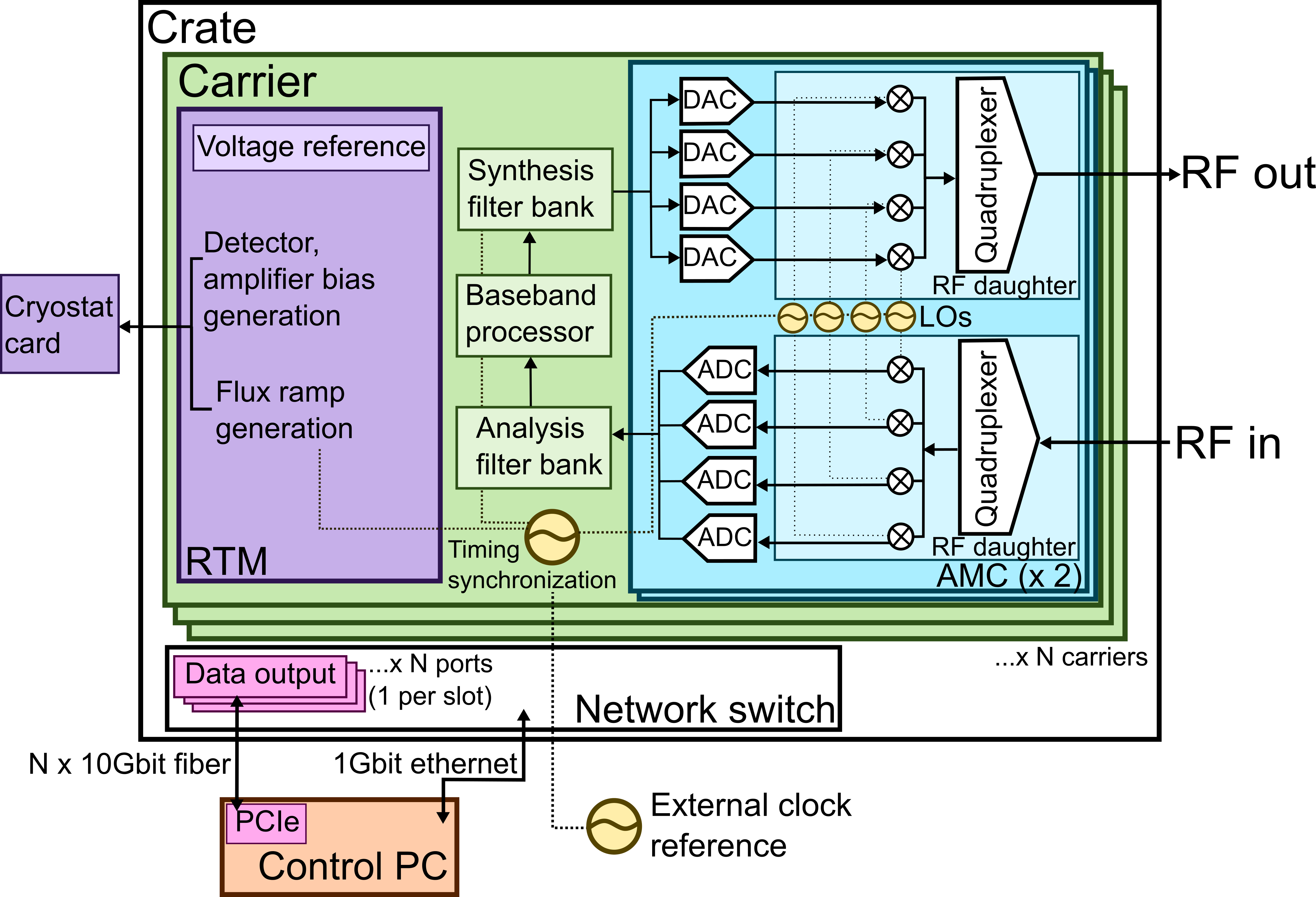}
	\caption{A sketch of a full SMuRF system. Each system consists 
      of an FPGA carrier board (green) that supports both the RF (blue)
      and DC (purple) systems. Multiple systems may be run in parallel 
      in a single crate. 
      The RF system (blue) consists of two Advanced Mezzanine Cards (AMCs), 
      each with a pair of RF daughter cards that interface between the RF 
      cabling to the cryostat and the ADC/DACs. The LOs used for up- and 
      downmixing signals between the baseband and RF are derived from the 
      main carrier but are generated on the AMCs. 
      The DC system (purple) provides the amplifier biases, TES biases, 
      and flux ramp generation on a Rear Transition Module (RTM). The 
      voltage sources are further conditioned in a cryostat card that 
      interfaces directly with the cryostat. 
      The main signal processing functions take place on the carrier, 
      and consist of a pair of polyphase filter banks for RF tone synthesis 
      and readback and a baseband processor for per-channel computations. 
      Several options exist for interfacing timing and data acquisition 
      with the system; displayed here are streaming to a PCIe card 
      installed in the control server and receiving an external clock 
      reference. 
      An ethernet connection between the the control server and network 
      switch provides firmware register access to all slots simultaneously 
      and may receive data streamed to disk for a small number of carriers. 
      Not pictured is the ATCA crate backplane, which distributes power
      and timing information and connects the network switch ports to
      the carriers in each slot.
    }
    \label{fig:full_sch}
\end{figure*}

The SMuRF electronics are built on the SLAC Common Platform, a
comprehensive FPGA-based firmware framework built on standardized hardware
designed specifically for high-performance systems.\cite{frisch16} 
The hardware components are designed for modular use with commercial 
Advanced Telecommunications Computing Architecture (ATCA) crates.\footnote{\url{
https://www.picmg.org/openstandards/advancedtca/}}
The hardware architecture is based on a Xilinx FPGA carrier board, which 
supports up to two double-wide dual-height mezzanine cards for 
analog and RF processing and a rear transition module for high 
speed links and further customization. 
A block diagram of the full SMuRF system is given in
Fig.~\ref{fig:full_sch}.

The system consists of an FPGA carrier card, which provides all
real-time signal processing. 
Mounted to the carrier are one or two Advanced Mezzanine Cards
(AMCs) which each consist of a base board and an RF daughter card. 
The base board contains the ADCs and DACs for the main signal path, clock
generation, and local oscillator (LO) generation.
RF daughter cards mounted to the AMC base boards contain the RF mixers, amplifiers, 
and bandpass filters. 
Opposite the AMCs on the carrier card is a Rear Transition Module 
(RTM) handling the low frequency signal generation, which is connected 
via a flexible multipin cable to a cryostat card that interfaces the
low frequency signals with the cryostat.
Additional peripherals such as the crate, timing, and networking are
required to run the system, but may be shared between multiple
systems.

Multiple parallel systems reside in a single commercial ATCA multislot 
crate, with each system occupying one slot.
Timing information is distributed through the crate backplane.
Typical crates contain $N+1$ slots, with $N$ carrier cards and one
slot dedicated to a network switch (Vadatech ATC807) that allows for
data streaming across multiple SMuRF systems.
A 1~Gbit ethernet connection between the ATCA network switch and the control 
server, commonly a Dell R440, provides firmware register access for every SMuRF 
carrier in the 7-slot ATCA crate, allowing for software control and data acquisition.
For large integrations, an additional PCIe card (Xilinx KCU1500) allows for streaming data from multiple carriers to the server.  
This specific hardware, including the PCIe card, Vadatech ATCA network switch, 
and Dell R440 computer server, have been qualified for operation in high-altitude 
environments as described in \S~\ref{sec:perfaltitude}.

Up to two AMCs fit into designated mezzanine 
slots on the FPGA, each with four up/downconverters corresponding to one of
eight 500~MHz wide bands.
The firmware framework is built on these 500~MHz-wide bands that can each accommodate 
processing and streaming of up to 416 simultaneous 2.4~MHz-wide channels. 
For each band, the tones are synthesized in the DACs and upconverted
to the appropriate GHz frequency range with a local oscillator.
Tones returning from the cryostat are split into bands and downmixed
with the same LO before being sent to the corresponding ADC.
The downmixed tones are digitized, channelized, and read into the
FPGA, where they are processed.
Each channel is typically associated with a unique resonance, thus creating 
a 1:1 mapping between readout channels and detectors. 
Thus, in total each SMuRF system contains the requisite hardware to read out
3328 resonator channels across 4~GHz of bandwidth, either in two parallel systems
reading out resonators in 2~GHz of bandwidth or in two serial systems
reading out resonators in across a 4~GHz range.

Each channelized tone is processed by the firmware to return data with desired 
levels of processing depending on the application.  
At the lowest level, the downconverted and channelized data can be streamed as the 
orthogonal I and Q components of the digital down conversion. 
Following calibration, these orthogonal components may be returned as the amplitude 
and phase components of the resonator voltage fluctuations, or an approximation of 
the probe tone frequency and resonator frequency error $\Delta f$. 
With this frequency error approximation, a feedback loop may be applied to minimize 
$\Delta f$, thus following the resonator minimum. 
This ``tone-tracking'' feedback enables the applied probe tone to ``follow'' the resonace 
dip as it is modulated by the flux ramp and/or incoming detector power. 

A more detailed discussion of the options for data processing is given in \S\ref{sec:firm}. 
At the highest level of processing, the SMuRF returns a time-ordered data stream of 
flux ramp demodulated detector data for saving to disk or passing on to downstream 
data acquisition. 

\section{\label{sec:hard}Hardware}

\begin{figure*}
\includegraphics[width=0.9\textwidth]{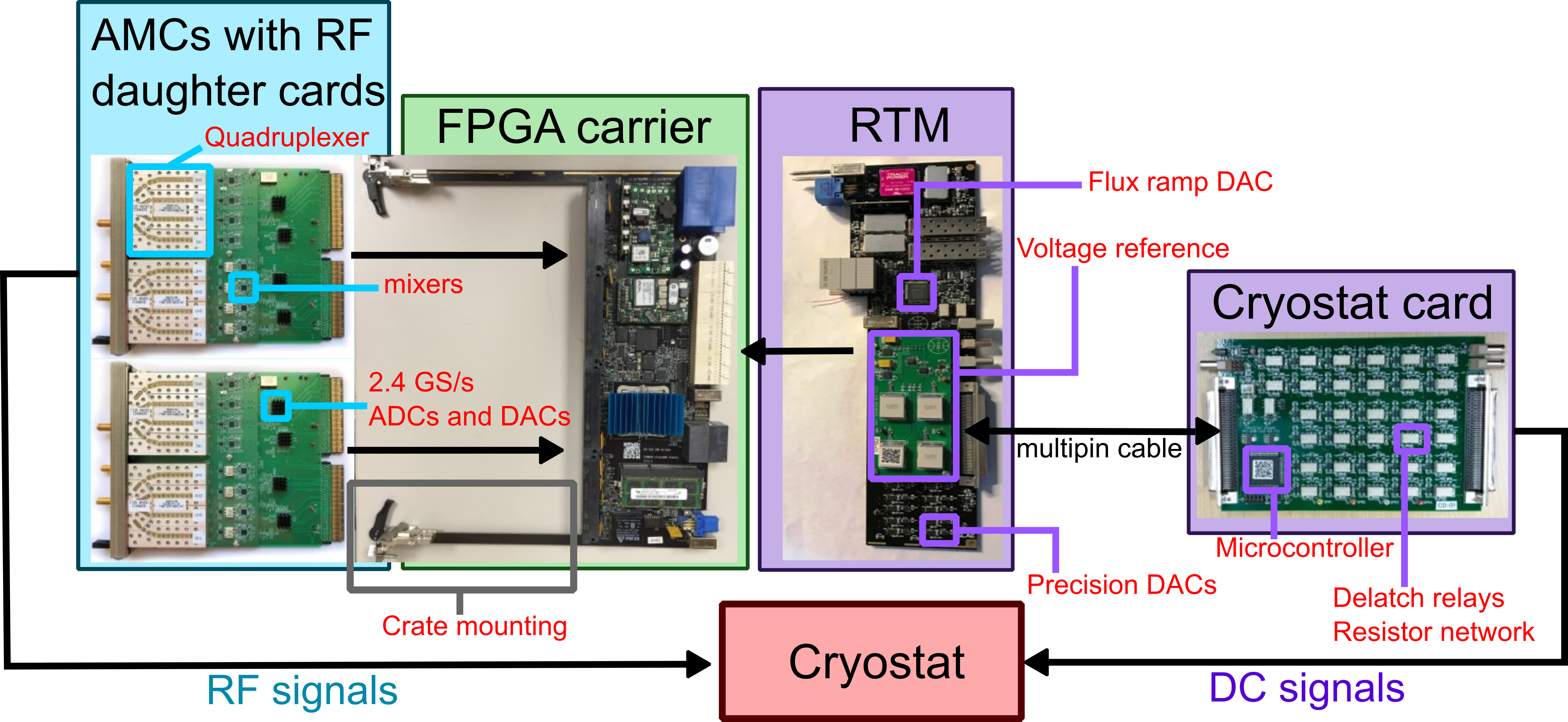}
\caption{\label{fig:hardware_pic}A picture of the hardware for one
  SMuRF system accommodating up to 3328 resonator channels across one
  or two RF chains, excluding peripherals such as the server, data
  streaming, and timing. The pictured contents excepting the cryostat
  card occupy one slot in a multi-slot ATCA crate. The AMCs with RF 
  daughter cards mount to the front of the carrier and present SMA 
  connectors for RF connection to the cryostat, while the RTM mounts 
  to the rear of the carrier and presents a multipin cable to a cryostat 
  card, which mounts directly to the cryostat. The RTM and cryostat 
  card combination supplies the low-frequency signal generation and 
  conditioning. 
}
\end{figure*}

A full picture of the hardware components is shown in Figure~\ref{fig:hardware_pic}.
The components displayed constitute the pieces of one SMuRF system,
which is capable of reading out up to 3328 channels in one rack-unit
of space, excluding peripherals that are shared between multiple
systems.
They are mounted in one slot of a multi-slot ATCA crate, and present 
SMA ports for RF connections from the front of the crate and a flexible 
multipin cable to a card mounted directly at the cryostat vacuum feedthrough 
at the back of the crate. 

\subsection{\label{sec:carrier}Carrier Card}

The FPGA carrier card (Xilinx XCKU15P Ultrascale+) handles the real
time signal processing, most notably the RF tone
generation/channelization and tone tracking.
The card provides interfaces for data, control, and timing.
An Intelligent Platform Management Interface (IPMI) provides a serial
control protocol for monitoring and power cycling.
A single-lane 2.5~Gbit uplink provides on board timing and a frequency
reference, while a 4-lane 10~Gbit/s ethernet interface to the ATCA
crate switch via the backplane handles all other communication,
including data transfer.
The AMC (see \S~\ref{sec:amcrf}) interface with the carrier consists
of eight 12.5~Gbit bi-directional JESD204b links to each AMC card for
ADC/DAC data, digital clocks, and low-speed serial peripheral
interfaces (SPI) for control and status readback.
The carrier interfaces digitally with the RTM via SPI link and provides
51.2~MHz clock (see \S~\ref{sec:rtm}).
The RTM additionally routes two optional external 2.5~Gbit external
inputs to the carrier, one for ethernet and the other for timing.

The firmware on the carrier card is programmed via VHDL and Xilinx System 
Generator for Digital Signal Processing and may be updated remotely. 
Further firmware details are addressed in \S~\ref{sec:firm}. 

Due to the large resource consumption of the signal processing
firmware and the need for many carriers to operate in close proximity
in high altitude environments, the carrier is robustly temperature 
controlled with large heatsinks, delidded regulators, and
high-velocity crate fans. 
The results of testing of these temperature control measures is 
described in \S~\ref{sec:perfaltitude}. 

\subsection{\label{sec:amcrf}Advanced Mezzanine Cards and RF Daughter Cards}

The Advanced Mezzanine Card (AMC) contains the ADCs and DACs for the
main RF signal path, clock generation system, and LO generation for the
RF daughter cards.
Each AMC consists of a base board interfaced with an RF daughter card supporting 
a total of 2~GHz of bandwidth at either 4-6~GHz or 6-8~GHz. 
The RF daughter card provides the up- and downmixing between the digital signal 
processing (DSP) band centered at 750~MHz and the RF frequencies of the microwave 
resonators. 
The current iteration of SMuRF has RF daughter cards for operation of either 
``low band'' (4-6 GHz) or ``high band'' (6-8 GHz) resonators. 
The daughter cards may be arbitrarily swapped between AMC base boards, which 
require only minimal modification between low and high band types. 
Each carrier card accommodates two AMCs of any combination of low band and high 
band types. 
The LO frequency may be modified via firmware register. 
Combined with RF hardware part swaps, this allows for for operation of SMuRF 
systems with cryogenic resonators at other frequency ranges and channel densities. 

\subsubsection{AMC Base Board}
\begin{figure}
\includegraphics[width=0.99\columnwidth]{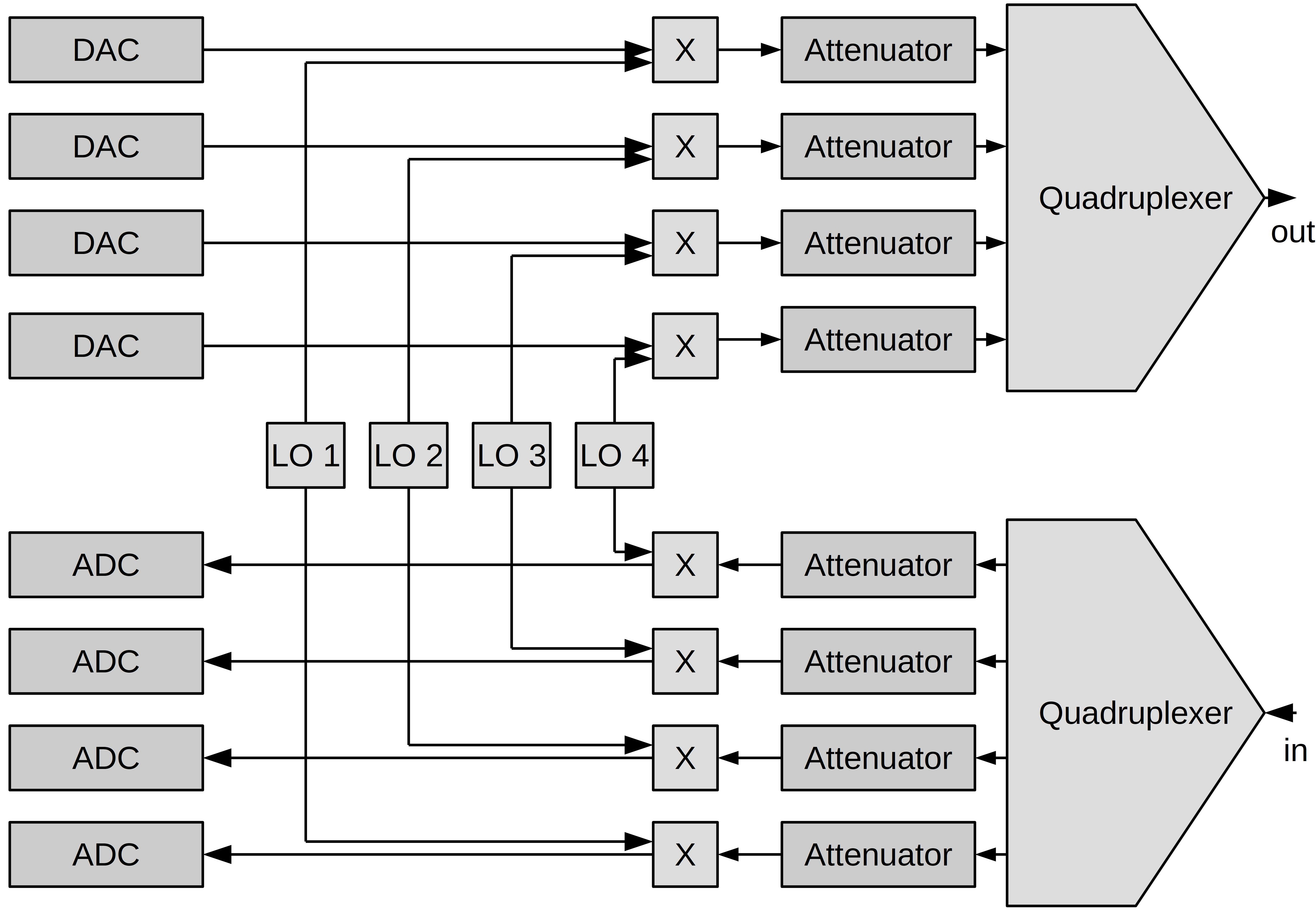}
\caption{\label{fig:cryodet_amc} Block diagram of a SMuRF Advanced
  Mezzanine Card (AMC), which is responsible for the RF signal
  input/output. RF signals enter the system between 4-8 GHz and are
  first split into 500MHz bands via a quadruplexer (bottom
  right). Each band is then processed in parallel through
  digitally controllable attenuators and mixers (denoted with X) each
  with an LO tuned to place the IF band in the 500-1000MHz range. The
  ADCs (bottom left) then sample the IF of each band at 2.4 GS/s. 
  The RF  output (up conversion) chain starts from the DACs (top left) and
  performs the same operations in reverse with the mixers fed by the
  same LOs as the down conversion. The separate bands are combined with 
  the quadruplexer, resulting in a comb of output between 4-8 GHz at 
  the RF output (upper right). 
}
\end{figure}

The AMC base board contains the hardware for interfacing the digital
logic on the carrier with the analog electronics.
A block diagram is shown in Fig.~\ref{fig:cryodet_amc}.

Each AMC baseboard features four 16-bit 2.4~GS/s ADCs (TI ADC32RF45) and four 16-bit
2.4~GS/s DACs (TI DAC38J84).
The SMuRF implements I/Q data separation digitally, using \emph{Digital DownConversion} (DDC) 
rather than use integrated analog I/Q combiners and separators.
In this scheme, rather than implementing separate I and Q hardware paths, a single 
high-speed ADC samples the full input signal, which is then separated 
into orthogonal components via multiplication with a directly synthesized intermediate 
frequency (IF), filtered, and downsampled to obtain the complex baseband signal. 
This relaxes the requirement on precisely matching the gain and
maintaining the phase offset between the I and Q components.
However, DDC requires higher speed DACs and ADCs to do the
I/Q combination and separation in firmware.

The DACs convert input complex data from the FPGA carrier card at 614.4~MS/s 
to a 500~MHz-wide filtered signal between 500~MHz and 1~GHz. 
The ADCs convert the unaliased input signal in the 500~MHz-1~GHz frequency 
range up to a complex data stream at 614.4~MS/s to pass to the carrier. 
The ADCs and DACs are run with 4x digital decimation and interpolation filters 
respectively to define the Nyquist band and simplify analog filtering requirements. 
Data bandwidth limits prevent the ADCs from transmitting wholly unprocessed ADC 
data and the DACs from processing arbitrary waveforms at the full data rate. 
The communication between the carrier and the ADCs and DACs is therefore routed 
through a 12~Gb/s custom, open source JESD protocol. 

Each DAC/ADC pair is associated with a unique local signal generator to provide 
the LO that the RF daughter card uses to upmix or downmix between the 500~MHz-1~GHz 
band and the desired GHz-frequency band. 
Each 500~MHz band has its own LO, which is offset by the ADC/DAC numerically-controlled 
oscillator (NCO) frequency of 2.4~MHz from the centers of 3.5~GHz, 4.0~GHz, 6.0~GHz, 
and 6.5~GHz for the low band, and 5.5~GHz, 6.0~GHz, 8.0~GHz, and 8.5~GHz for the 
high band.
We use the lower sidebands for the first two LOs and the upper sidebands for the 
second two such that the LOs fall outside the resonator bandwidth when operating 
AMCs of the same type (both low or both high band). 
These LOs are generated via a fractional-N locked phase-locked loop IC (Analog 
Devices ADF5355) with integrated low-noise voltage-controlled oscillators (VCOs).

Finally, the AMC card contains a clock generation system (TI LMK04828) for the 
digitizer and associated JESD204b links. 
This clock, as well as all other SMuRF system clocks, are ultimately
derived from a common timing reference (Crystek CVHD-950).

Since the LO is used to generate all the tones in its respective 500~MHz band, 
its noise profile is critical to the overall noise performance.
Other studies find that the noise profile of the LO can be seen in 
resonator channel noise spectra at high flux ramp rates, which does not
impact typical science data but may constrain the maximum operable 
flux ramp rate.\cite{ltd21} 
The noise profile of the LO is set by the noise of the reference clock source, 
the charge pump used for loop gain adjustment, and the VCO. 
The phase noise of the clock reference source the LOs are locked to sets the 
close-in phase noise (below $\sim$100~Hz).
An external clock reference can be passed in and locked to the onboard clock 
either via a SMA input on the AMC front panel or through a timing distribution 
system on the ATCA crate backplane.
If no external source is provided, the 122.88~MHz is derived from the FPGA. 

Above 100~Hz the phase noise of the LO is set by the parameters of the charge 
pump phase-locked loop (CP-PLL) which is used to create a flat phase noise profile 
out to \Order(100~kHz). 
The noise level in this flat portion of the phase noise profile can be adjusted 
slightly by trading off bandwidth of the CP-PLL. 
Pushing to much higher bandwidth above $\sim$ 100s of kHz leads to instability 
in the CP-PLL. 
The phase noise after the CP-PLL rolloff is set by the intrinsic phase noise 
of the VCOs that generate the LO, which ultimately limit the phase noise level. 
For applications that require lower system phase noise, the VCOs can be exchanged with lower phase noise, fixed-frequency crystal oscillators.
These were not used in the current design because flexibility of LO
frequency while achieving the required white noise specification 
was favored over further improvements in phase noise.

\subsubsection{AMC RF Daughter Cards}

The AMC RF daughter cards mounted to the AMC base boards provide the
RF frequency tone generation and readback.
The RF mixers (Analog Devices LTC5548) that provide the up- and
downmixing between between the DAC/ADC baseband from the AMC base board 
and the LOs are contained on the RF daughter cards.
Since the RF mixer typically sets the linearity of the system, the choice of 
mixer here is critical, as is the attenuation of DAC output power prior to 
passing to the mixer. 

The RF daughter cards exist in both low band (4-6~GHz) and high band 
(6-8~GHz) variants. 
Quadruplexers (Lark Engineering SMCQ819-110 and SMCQ819-111) interface from 
the mixers on the four ADC/DAC pairs to a single input/output coaxial pair 
that connect with the cryostat or other test system via SMA connectors. 
Cavity filters within the quadruplexers define overlapping bandpass filters 
that split the RF frequency band into each 500~MHz band that is synthesized 
or channelized in firmware. 
Since the ADCs and DACs are clocked at 2.4~GS/s, the Nyquist frequency for the 
alias band is at 1.2~GHz.

\begin{figure}
\includegraphics[width=0.99\columnwidth]{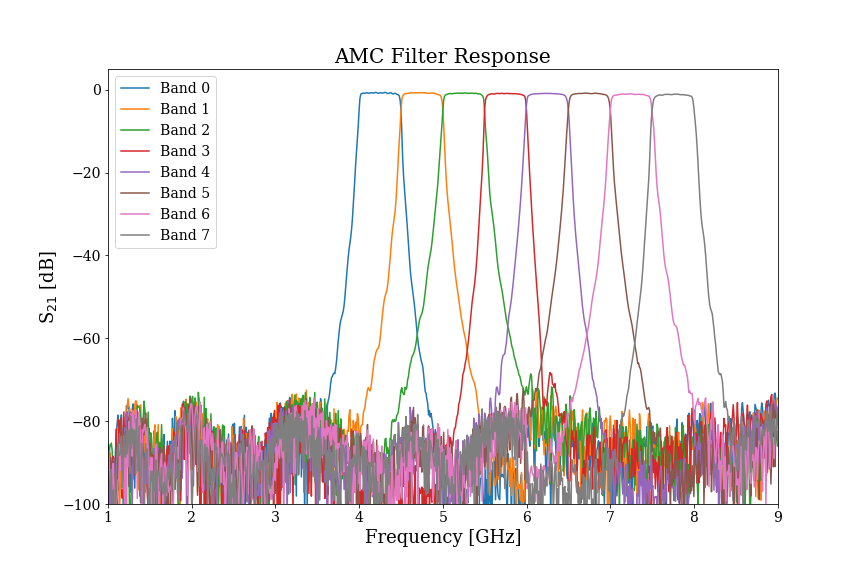}
\caption{\label{fig:amcfilts} The response of the AMC band-defining filters 
for each of the 500~MHz wide bands across the full 4-8~GHz SMuRF band. Band 0 
corresponds to the 4-4.5~GHz band, and the bands count upwards in 500~MHz 
increments from there. }
\end{figure}

The AMC $\mathrm{S_{21}}$ response for each 500~MHz band is shown in 
Figure~\ref{fig:amcfilts}. 
We note in particular that the maximal passband of the filters is not quite 
500~MHz wide to guarantee sufficient rejection out of band, set at 60~dB 
isolation between bands as described in \S~\ref{subsec:ct}. 
The portion of the passband that is flat within -2~dB is about 460~MHz wide, 
while the edges of each band are about 3~dB down and isolation is \textgreater 
~65~dB between adjacent bands. 
The RF daughter cards additionally include programmable attenuators on both 
the RF input and output for each 500 MHz band to allow for flattening the
 RF signal levels over the full SMuRF RF bandwidth. 

Since the ADCs and DACs have optimal noise performance at a fixed output/input 
power level, respectively, adjusting power levels at the resonator is most easily 
accomplished via some combination of external attenuation/amplification and these 
programmable attenuators. 
With existing default settings, the optimal output/input power levels are about 
-33~dBm per tone. 
Individual tone powers are adjustable in firmware in coarse (3~dB) steps, while 
the programmable attenuators act on each 500~MHz band defined by one ADC/DAC pair 
and can be tuned in finer (0.5~dB) steps. 
The size of firmware steps can be modified easily in firmware, but the present 
combination of coarse and fine steps has been sufficient for user needs. 

\subsection{\label{sec:rtm}Rear Transition Module}

The rear transition module (RTM) generates the following low frequency
signals for the SMuRF system: (1) Detector biases, (2) RF amplifier biases, and 
(3) flux ramp. 
It additionally provides control and clocking to the cryostat card. 

The RTM acquires its sequencing and interfaces via a complex programmable 
logic device (CPLD), which runs a simple, compact firmware programmed in VHDL.
The CPLD firmware cannot be updated remotely. 

The DC TES biases and cryogenic amplifier biases are provided by 33 low-noise 
18-bit DACs (Analog Devices AD5780). 
To reduce low-frequency noise, the TES bias DACs are typically operated 
differentially, though they can be driven single-ended with no modifications 
to the SMuRF hardware or firmware to double the number of independent TES bias 
lines.

As discussed in \S\ref{sec:integration}, for applications requiring high RF 
linearity it may be desirable to cascade two cryogenic amplifiers rather than 
use a single LNA. 
Thus, the RTM allocates eight DACs to biasing up to four total RF amplifiers, 
typically two amplifiers placed at 4K and 50K for two separate RF chains. 
Each amplifier is allocated two DACs for the gate and drain voltages, though some 
models require only a single bias line. 
Additional amplifier bias lines can be added at the expense of TES detector bias 
lines with no modification to the RTM hardware or firmware, if desired. 
These DACs, referred to as ``slow'', can generate arbitrary waveforms
up to about 1~kHz for calibration and testing purposes.

An additional 50~MS/s 16-bit differential current DAC (Analog Devices
LTC1668) drives a differential pair that serves as the flux ramp for
\umux systems.
Since the flux ramp is required to be much faster than the detector
and amplifier biases and operates at lower full scale voltages, it is 
higher speed and therefore higher noise than the DC bias DACs. 
The flux ramp is implemented as a sawtooth function, although other
waveforms are possible.
It is generated by the RTM CPLD with control and triggering from the
carrier card FPGA.
The RTM finally provides DC power and a low speed SPI serial link for
communication to the cryostat card (see \S~\ref{sec:cryocard}).

A daughter card on the RTM consists of a voltage reference providing
buffered $\pm 10$V reference signals to the precision DACs.
The card uses two commercial voltage references (Analog Devices
LTC6655) in parallel to reduce noise, and low-pass filters with low
acoustic sensivity capacitors (Rubycon 35MU106MD35750) to
mitigate vibrational pickup.
A daughter card implementation for the voltage reference provides the 
flexibility to easily and cheaply upgrade the performance of the voltage 
reference if required for a particular application.

The RTM presents a 100-pin SCSI connector containing the low-impedance
flux ramp, TES bias, amplifier bias, DC power supplies, and digital
communication to the cryostat card.
The SPI communication link between the RTM and the cryostat card is
default quiet to reduce noise during operation, with no active
polling.

Two monitor ports with LEMO connectors are available laboratory diagnostics.
First, the flux ramp monitor port outputs a single-ended 
50~\Ohm-impedance signal for monitoring the flux ramp output. 
This line is not guaranteed to have the requisite noise stability or 
bandwidth for properly driving the flux ramp.
Second, the trigger output provides a low voltage TTL signal asserted at the
start of each flux ramp period, typically at the resets for the
sawtooth signal.

An SFP 1~Gbit ethernet port may be used for data streaming of a single SMuRF system 
without the need for interfacing with the rest of the ATCA crate. 
An additional SFP port receives timing data that may be used for synchronizing 
multiple SMuRF systems together, as discussed in \S~\ref{sec:misc}. 

\subsection{\label{sec:cryocard}Cryostat Card}

The cryostat card interfaces signals from the RTM directly with the
cryostat.
It conditions the TES bias and flux ramp signals from the low-impedance
voltage sources at the output of the RTM to filtered, high-impedance
current sources.
Since the bias lines are particularly susceptible to noise pickup
after converting to current sources, the cryostat card was designed to
be placed as close to the cryostat as possible, with a separate card
and enclosure from the main RTM to allow for shielding and thermal
regulation.
The connector on the cryostat side is chosen to match the feedthrough of 
the desired application. 

The cryostat card additionally provides regulated power supplies and conditions 
gate bias voltages from the RTM DACs to bias the cryogenic RF amplifiers.
The card is controlled and monitored via an on-board microcontroller
(Microchip PIC32MX795 32-bit microcontroller) and interfaces to the
rest of the SMuRF system via an SPI serial link to the RTM CPLD.

The detector bias is a differential output voltage using
two opposing $\pm10$V RTM DACs.
For each detector bias line, mechanically latching relays on the
cryostat card (Axion 5-1462037-4) controlled by the microcontroller
allow for switching between a low-pass filtered, low-current bias
circuit (``low current mode'') and a higher current, minimally
filtered ($\sim$ 100~kHz bandwidth) circuit (``high current mode'').
The relays are mechanically latching to reduce power dissipation that
could couple to temperature variation in the board, meaning the state 
cannot be queried during operation. 
The default state is in low current mode.

In low current mode, added parallel capacitance low pass filters the
detector bias outputs with a frequency cutoff $f_\mathrm{3dB} \sim
8~\mathrm{Hz}$.
The output current levels in both low and high current modes are set
with inline series resistors, typically in the 1-50~k\Ohm range
depending on the maximum current desired, providing a high impedance
source.
The maximum current is thus determined by the resistor selection,
which is set by individual detector operating parameters, up to op-amp
limits of $\sim 10$~mA.
This high current mode is typically used for delatching superconducting detectors, 
biasing detectors with higher operating resistances for calibration purposes, 
or for high bandwidth measurements.

As discussed in \S~\ref{sec:reqdes}, we set stringent requirements on the low 
frequency performance of signals provided by the cryostat card. 
It is expected that temperature variation of electronic components drives a 
large fraction of the variation in the low frequency noise, particularly the 
temperature variation in the RTM voltage reference and cryostat card resistors. 
This sets a requirement of <$\sim$20ppm/\degr C on the temperature coupling 
coefficient of these resistors. 
The cryostat card resistors were thus chosen to be guaranteed to 
$\pm 10$~ppm/\degr C (TE Connectivity RQ73 series). 
The capacitors on the cryostat card were similarly selected for low frequency 
noise performance and low acoustic sensitivity (Rubycon).

The 4K and 50K amplifier gate bias voltages are each driven single ended via 
an RTM DAC with $20~\mathrm{\mu F}$ bypass capacitors on the cryostat card 
for filtering.
A resistor divider on the cryostat card allows for limiting the output gate 
voltages to less than the maximum amplifier ratings if required.
The drain currents may be read back to allow for bias adjustment using an 
inline resistor on the input to the regulator, which is amplified by an 
instrumentation amp and read back by ADCs on the microcontroller.
The flux ramp and amplifier bias cryostat card circuits have jumpers that allow
the user to manually select between sourcing from the RTM and inserting signals
from an external source via a 2-pin differential LEMO connector.

All output voltages, including the RF amplifier drains, are disabled 
by default on   power-on, and must be explicitly enabled.  
This is a requirement for many cryogenic RF amplifiers, which can 
draw large currents and generate substantial heat loads if the drain 
voltage is enabled before the gate voltage is set.  
For the RF amplifier drain voltages, which are generated on the cryostat 
card using adjustable linear voltage regulators, this is accomplished by 
toggling the enable pin of the linear regulators using the cryostat card PIC.

The cryostat card provides an additional DC line at +12V with minimal
filtering for powering an optional warm follow-on RF amplifier, which
can be used to optimize the cryostat RF output signal power levels
into the AMC RF ADCs.
For general diagnostic purposes, it can report its temperature using an 
on-board TI TMP236 temperature sensor over the SPI bus.

The flux ramp is a differential output current source drive, converted
from the RTM low-impedance voltage source via a passive resistor H-pad
attenuator to allow impedance matching if required by the application.
A mechanically latching relay allows for switching between AC and
DC-coupled modes via large capacitors. 
In AC mode, large acoustic sensitive capacitors (Rubycon) are used 
to mitigate against vibrational coupling. 
Operating in AC mode high-pass filters the differential flux ramp
drive with an $f_\mathrm{3dB}\sim 80$~Hz, attenuating 60~Hz power line noise 
and low frequency noise pickup in the cables between the RTM and cryostat.  
Operating in DC-coupled option enables low-frequency and fixed-current flux ramp biasing 
as required, such as when taking \umux resonator frequency versus flux ramp bias 
curves or other device characterization data. 

\subsection{\label{pcie}PCI Express Card}
For large integrations, a Xilinx KCU1500 data acquisition PCI express (PCIe) 
card using a custom firmware allows for multiple carriers' worth 
of data to be streamed to a server, with each slot outputting to a 
10~Gbit fiber link on the network switch.\cite{kcu1500}
The PCIe firmware implementation is designed for a 7-slot ATCA crate configuration, 
with the PCIe card interfacing with up to six SMuRF carriers through the Vadatech 
switch via dedicated ethernet links to an eight lane PCIe 3.0 bus in a server 
computer (commonly a Dell R440).
The dedicated ethernet link for each SMuRF carrier is implemented with 10~Gbit 
fibers which connect from the ATCA ethernet switch to two Quad Small Form-factor 
Pluggable (QSFP) transceivers in the KCU1500. 
The PCIe card presents two x8 interfaces bifurcated to a x16 edge connector to the 
server. 
During data taking, each SMuRF carrier transfers data frames at the flux ramp rate 
to the PCIe card, where the frame stream is received and processed further in software. 
Common software processing steps include packaging the data with an application-specific 
header, which can include timing information, and optionally filtering and downsampling 
the data streams before writing the data to disk.

\section{\label{sec:firm}Firmware}

\begin{figure*}
\includegraphics[width=0.9\textwidth]{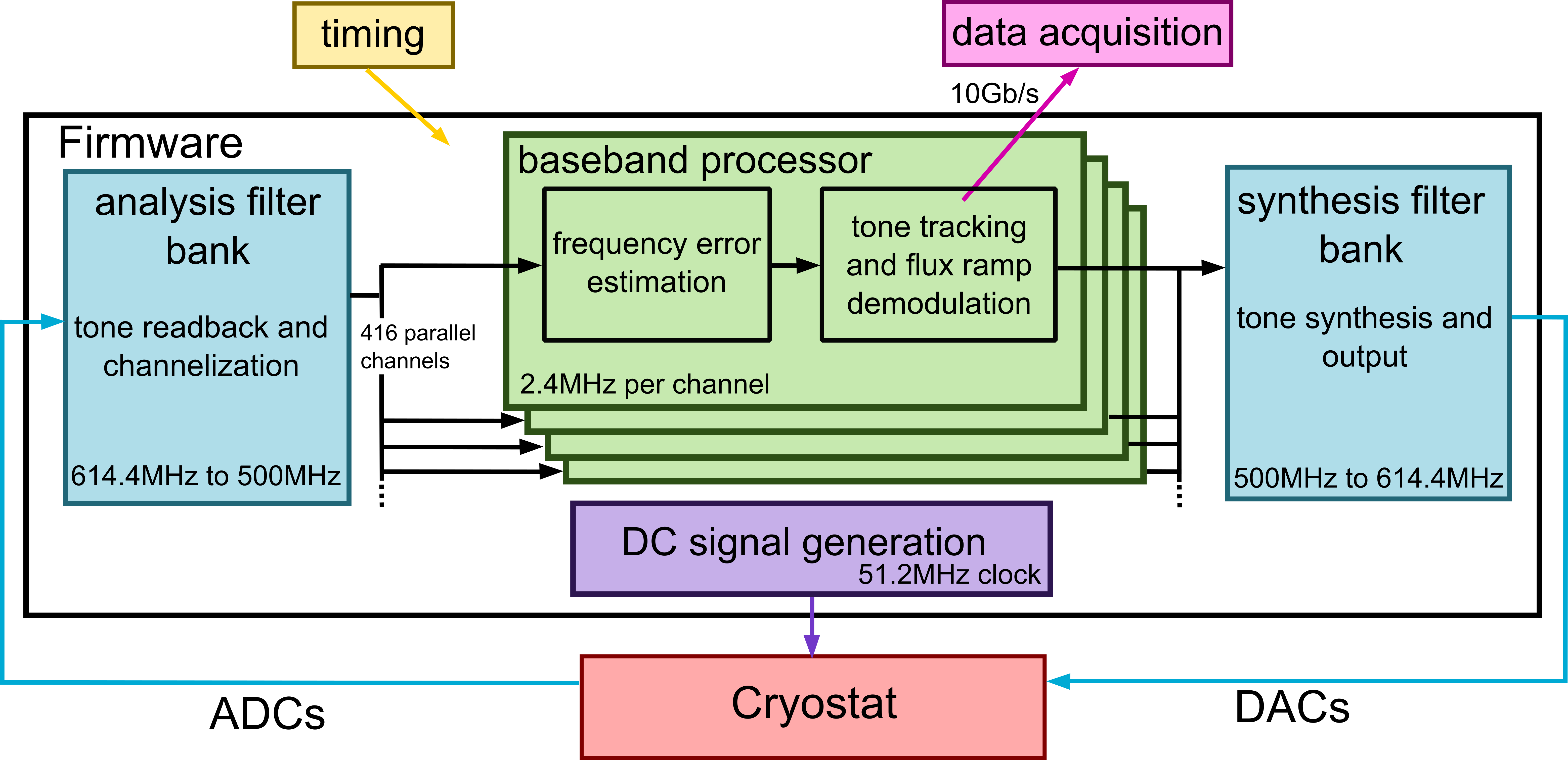}
\caption{\label{fig:firmware_overview} Block diagram of the main components of 
the SMuRF firmware. Some miscellaneous tasks are not shown for simplicity.
The blue boxes represent a pair of polyphase filterbanks that provide the synthesis 
and readback of the RF tones used to probe the resonators. 
The green box contains the baseband processor, which acts on each channel separately 
to perform the calibration and frequency error estimation, then the tone-tracking and 
flux ramp demodulation. 
The filter banks and baseband processor act on hardware-defined 500~MHz-wide bands. 
One SMuRF FPGA carrier card supports up to 8 bands, allowing for a total of 3328 channels 
per board. 
Auxiliary firmware functions such as timing synchronization, data framing, and 
low-frequency signal generation are represented by the yellow, pink, and purple 
boxes respetively. }
\end{figure*}

The FPGA carrier card firmware is responsible for SMuRF computation including RF tone 
generation, tone tracking, flux ramp demodulation, and data framing. 
A component block diagram is given in Figure~\ref{fig:firmware_overview}. 
The SMuRF firmware components provide the following functionality:
\begin{itemize}
    \item \textbf{Tone Generation and Channelization:} A pair of polyphase filter banks 
    provide synthesis and analysis of the RF tones used to probe the resonators. The filter 
    banks are show as the blue boxes in Figure~\ref{fig:firmware_overview}.
	\item \textbf{Frequency error estimation:} SMuRF uses an affine transform to translate 
	the received complex response (S$_{21}$) into a resonator frequency error estimate. 
	As this occurs on a per-channel basis, it is contained in the baseband processor (green 
	box in Figure~\ref{fig:firmware_overview}). 
    \item \textbf{Tone tracking and flux ramp demodulation:}  SMuRF uses a feedback loop 
    to update the probe tone frequency, attempting to minimize its estimate of the 
    frequency error.  The feedback loop parameterizes the flux ramp modulation and 
    also outputs the demodulated detector signal. This is also part of the baseband processor. 
    \item \textbf{Miscellaneous:} The firmware handles miscellaneous timing and waveform 
    synthesis tasks, such as flux ramp timing, arbitrary detector bias DAC waveform 
    generation, data streaming, and synchronization between multiple SMuRF systems. These 
    are shown heuristically with the timing (yellow), DC signal generation (purple), and 
    DAQ (pink) boxes in Figure~\ref{fig:firmware_overview}. 
\end{itemize}
Intermediate data outputs with increasing levels of signal processing are available to examine for system characterization and debugging. 

\subsection{\label{sec:pfb}Tone Generation and Channelization}

\begin{figure}
\includegraphics[width=0.9\columnwidth]{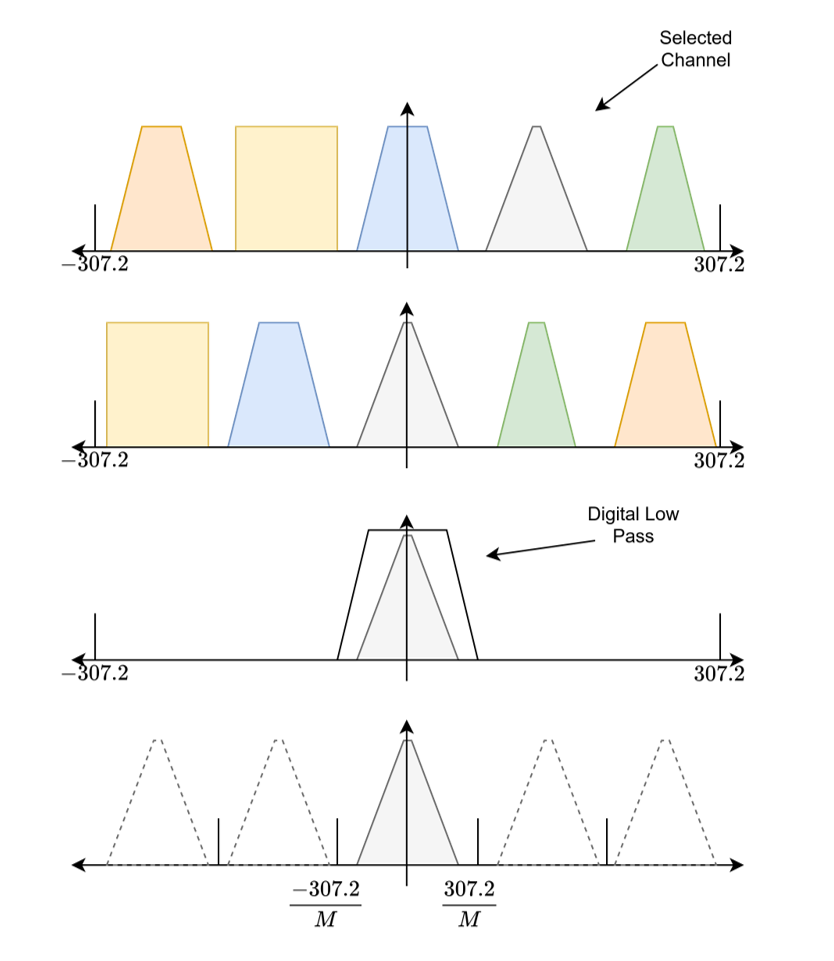}
\caption{\label{fig:downconverter} Conceptual diagram of the downconversion process 
for a single SMuRF channel. Heuristically, the process works as follows: \\
(1) The incoming data $x[n]$ consists of $N$ channels separated in frequency space 
across the 614.4~MHz-wide output of the ADC, centered each at frequency $\theta_k$ 
for $k=1,2,\ldots,N$. \\
(2) Each channel is centered by multiplying with the inverse of the channel center 
frequency $e^{-j\theta_k n}$. \\
(3) A digital low-pass prototype filter is applied. \\
(4) We downsample by $M=256$ to arrive at a 2.4~MHz-wide single-channel output. \\
This process is implemented in parallel for $N=512$ channels via the analysis 
filter bank. The similar upconversion process occurs via the synthesis filter 
bank in the reverse order.}
\end{figure}

The SMuRF FPGA synthesizes and analyzes data from each of 500~MHz bands 
corresponding each to one RF ADC/DAC pair separately. 
The RF ADCs and DACs transform the IF band from 500~MHz-1~GHz to and
from the complex baseband of $\pm$307.2~MHz, defined by the
ADC sample rate of 2.4~GS/s and divide-by-4 complex
decimation filter.
From here, the data is downconverted from a 614.4~MHz-wide frequency
band to individual channels to be processed.
A conceptual sketch for the single channel downconversion chain is
given in Fig.~\ref{fig:downconverter}.  The upconversion works
similarly, but in reverse.

\begin{figure}
\includegraphics[width=0.99\columnwidth]{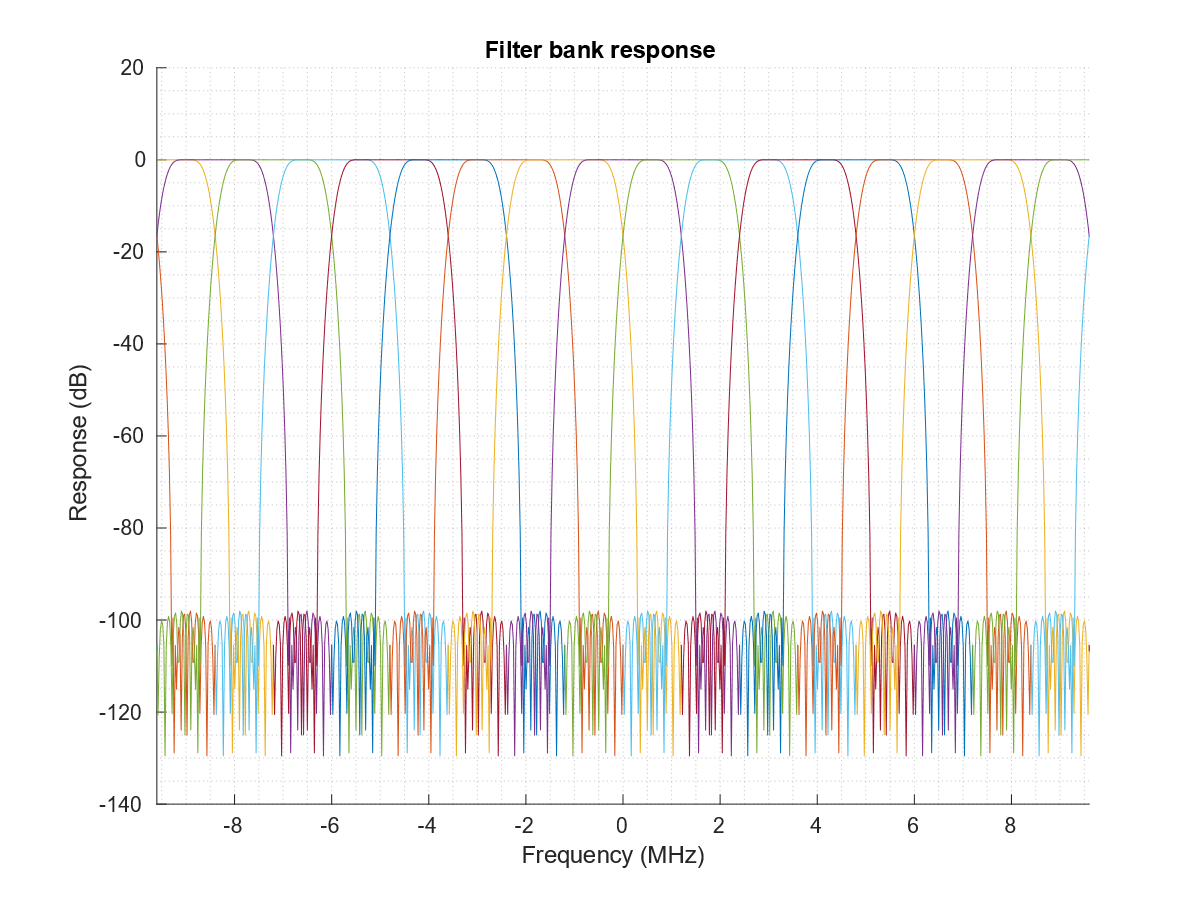}
\caption{\label{fig:filterbankresp} A zoomed-in section of the filter
  bank response for 16 interleaved 2.4~MHz wide channelization
  bins. Note that the channels are oversampled by 2 such that the bins
  are overlapping with each other. The filter bank converts between
  614.4~MHz of bandwidth and 512 time multiplexed bins that are each
  2.4~MHz wide.}
\end{figure}

A polyphase filter bank performs the channelization, converting from
each 614.4~MHz I/Q stream to 512 interleaved 2.4~MHz-wide bins that
are time multiplexed.
The filter bank response is shown in Fig.~\ref{fig:filterbankresp}.
The filter bank oversamples by 2 such that each resonance falls into
multiple bins; two resonances which fall in the same bin may be
assigned to an overlapping bin such that both resonators are
processed.
Thus, a given 1.2~MHz frequency band may not contain more than 3
resonances such that every resonance can be assigned to its own bin.
The firmware is capable of supporting processing multiple channels per bin 
if desired, but currently uses only one channel per bin. 
Placing a single channel per bin introduces additional latency due to the 
lack of parallelization, but reduces FPGA resource consumption. 
NIST CMB-style \umux resonators aim for at least 1~MHz
spacing between \umux resonances for crosstalk
considerations; thus the width of the filter bank channels does not 
significantly contribute to readout yield degradation.\cite{mates19}

The data from each bin is filtered with a 16 tap digital low-pass
filter, resulting in a 4096 tap FIR filter (since there are 512
resulting channels, oversampled by 2).
The filter bank achieves about 100~dB of rejection for out of band
response and adds a group delay that is approximately constant across
the 2.4~MHz-wide band.

This filter sets the majority of the processing delay due to SMuRF, on
the order of 6~$\mathrm{\mu s}$ out of a total signal processing delay
of about 8.8~$\mathrm{\mu s}$.
For comparison, the roundtrip cable delay through a typical CMB
receiver cryostat may be about 100~$\mathrm{ns}$, and the additional
delay from the ADCs and DACs another $\sim 0.5~\mathrm{\mu s}$.

The FFT is performed with a streaming Radix $2^2$ architecture 
to convert between the frequency and time domains.

The tone frequencies are defined with a 24-bit input to a 
\emph{direct digital synthesis }(DDS) system. 
This results in a frequency resolution of 0.1431~Hz for probe tones, 
which is essentially continuous for practical purposes. 
A 10-bit look-up table and Taylor series correction generates the probe 
tone per channel, which is amplified to the desired output tone power and 
converted back to the full frequency-multiplexed bandwidth via the synthesis 
filter bank. 
The DDS system achieves greater than 100 dB spur-free dynamic range (SFDR) 
over the synthesis filter bank per-channel bandwidth of 2.4~MHz. 
The DDS continuous synthesis allows for fine frequency resolution and fast, 
phase-continuous frequency switching necessary for continuously updating the 
tone frequency with tone tracking, as discussed in \S~\ref{sec:trackingdemod}. 

The 512 time-multiplexed channels output from the polyphase filter bank are 
processed in two parallel streams by the baseband processor, which performs 
the per-channel calibration and feedback operations discussed in \S~\ref{sec:eta} 
and \ref{sec:trackingdemod}. 
While the filter bank channelizes the full 614.4~MHz bandwidth, we only use the 
center 500~MHz per AMC as defined by hardware filters. 
To reduce power consumption on the FPGA, the edge channels are dropped from 
processing, resulting in 499.2~MHz of bandwidth covering 208 time multiplexed 
channels per baseband processor stream for a total of 416 channels available 
per 500~MHz band. 
While dropping the edge channels adds the complication of converting
the 614.4~MHz stream from the ADC to a 499.2~MHz stream for
processing, the slower clock speed is critical to fitting the baseband
processor computations within the resource limits of the FPGA.

For each channel, the ADC output is digitally downconverted with the output 
from the DDS to center it about 0~Hz.
At this point, the data may be 
optionally output as per-channel orthogonal I and Q streams for offline test and 
debugging purposes.
Application of calibration (as discussed in \S~\ref{sec:eta})
transforms this data into tone frequency and frequency error between
the output tone and the resonator frequency.
Finally, an adaptive loop filter configured with feedback parameters (as
discussed in \S~\ref{sec:trackingdemod}) outputs flux ramp demodulated
data at the flux ramp frame rate, and feeds back to the DDS to update
the output tone frequency during tracking.

After interacting with the baseband processor, the
time-multiplexed processed channels are padded from 499.2~MHz back to
the 614.4~MHz bandwidth and synthesized into a 614.4~MHz I/Q stream
via the synthesis filter bank.
A user-defined delay may be added to correct for any system delays due to
cabling, processing, etc. before being passed to the DAC.
The relevant frequency domains of the channelization and synthesis
process is given in the main RF loop of
Figure~\ref{fig:firmware_overview}.

\subsection{\label{sec:eta}Frequency Error Estimation}

\begin{figure*}
\includegraphics[width=0.99\textwidth]{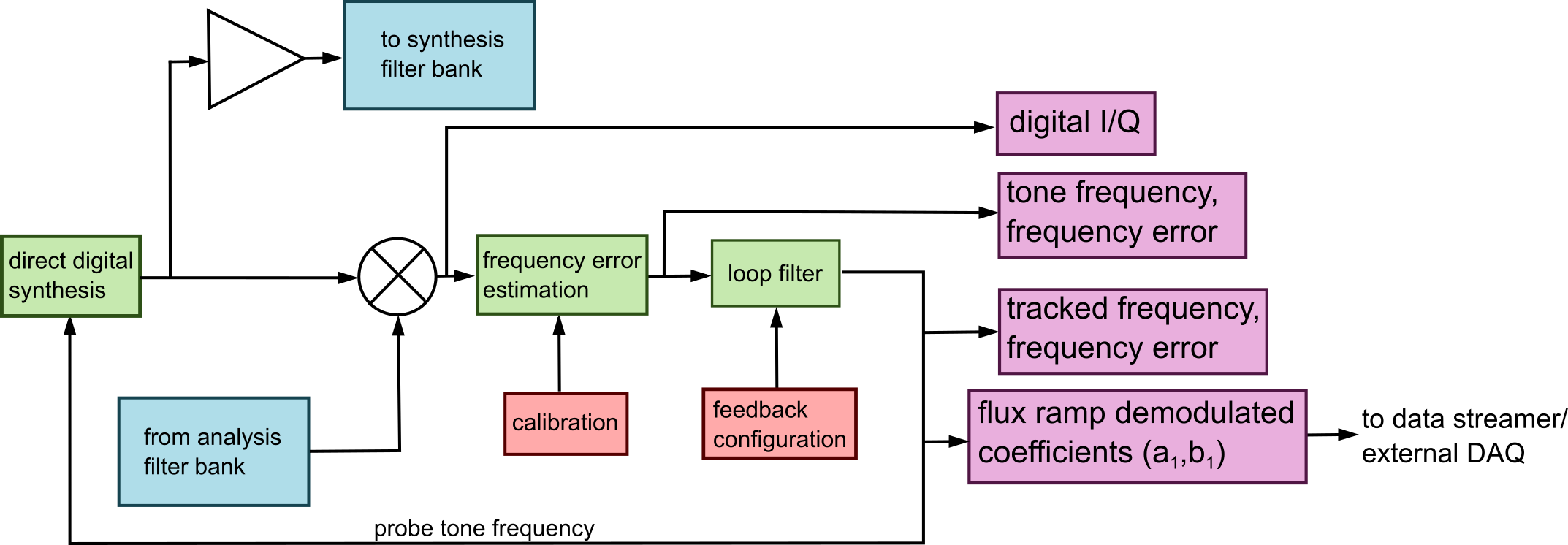}
\caption{\label{fig:baseband}The per-channel baseband processor. This
  acts on each of the channels serially, clocked at 499.2~MHz. Various
  intermediate points are available to examine for debugging and more 
  detailed characterization, or the final demodulated data is passed to 
  the data streamer for packaging and writing to disk or to external 
  DAQ. External inputs are shown in red, FPGA computations are shown in 
  green, interactions with the filter banks are given in blue, and options
  for data outputs are in pink.}
\end{figure*}

Following the tone synthesis and analysis, the SMuRF electronics can
be thought of as a system that is capable of generating tones and
reading back the complex response as a channelized stream of orthogonal 
in-phase and quadrature components, typically denoted I and Q respectively.
This I/Q stream may then be processed by the baseband processor, which acts on 
each channel serially. 
A conceptual diagram of the baseband processor is given in Fig.~\ref{fig:baseband}.

For \umux systems, we are typically interested in modulation of the resonance 
frequency due to incoming power on the detector. 
We thus seek to transform our complex digital data to an estimate of
resonator frequency shifts.
We transform shifts of the resonance frequency into shifts entirely in
the phase ($Q$) direction of the resonator response via a rotation and
scaling of the resonator's complex response, allowing us to ignore
changes in the amplitude direction ($I$).
Equivalently, we rotate the resonator circle in the complex plane such
that it is oriented as given in Figure~\ref{fig:excirc}, and
approximate small movements near the resonance frequency as occurring
along a single axis.

The resonance frequency is estimated as the point of minimum amplitude
response, which is true for an ideal resonator and a good
approximation for typical NIST CMB-style \umux resonators.
Resonance frequency estimation is performed in software; thus other
metrics may be supported by the SMuRF system as various applications
require.

The resonator has a response in the complex plane corresponding to the
real and imaginary components of the S$_{21}$ response.
This response may be rotated and scaled by a complex number $\eta$
such that for small changes in resonance frequency, the change in
S$_{21}$ is entirely in a single quadrature.

For $(I_\pm,Q_\pm)$ the complex transmission measured at frequency offset 
$\pm\Delta f$ from the resonance frequency, we estimate $\eta$ as 
\begin{equation}
\label{eq:eta}
\eta = \frac{2\Delta f}{(I_+ - I_-) + i(Q_+ - Q_-)}.
\end{equation}
\noindent We see that the denominator gives an angle in the complex
plane by which to rotate the resonance circle, while the overall
magnitude is set by the sharpness of the resonance.
Thus, $\eta$ has the effect of rotating and scaling the resonance
circle such that the real and complex axes have physically meaningful 
units and with the S$_{21}^\mathrm{min}$ rotated to be 
parallel to the imaginary axis. 

\begin{figure*}
\includegraphics[width=0.99\textwidth]{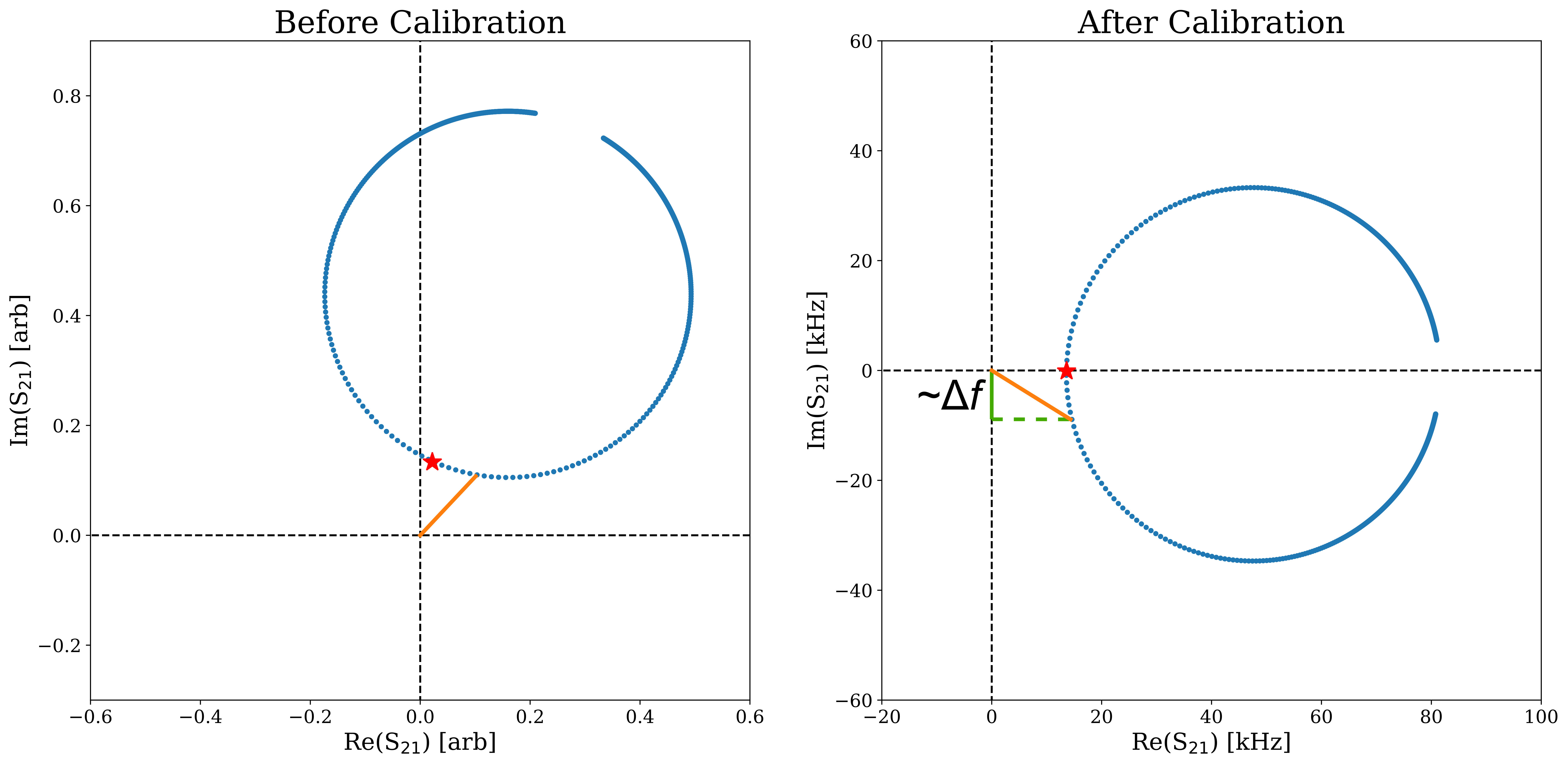}
\caption{\label{fig:etacal} A sketch of the calibration and frequency error estimation. 
[Left] An uncalibrated resonator generically has some response in the complex 
plane that may be arbitrarily rotated and scaled. 
A probe tone (orange) is initially tuned to the point of minimum transmission 
S$_{21}^\text{min}$ (red), but as the resonator is modulated by flux the 
S$_{21}^\text{min}$ will shift away from the probe tone. 
[Right] Following calibration by $\eta$ as defined in Eq.~\ref{eq:eta}, the 
resonance circle is scaled such that the axes are in physically meaningful 
units and the small changes in the resonance frequency correspond to movements 
in only one axis. The frequency error $\Delta f$ between the probe tone and the 
new resonance frequency is estimated as the projection (green) of the probe tone 
complex response onto a single axis, corresponding to Eq.~\ref{eq:deltafeta}. 
}
\end{figure*}

With this calibration, the frequency shift may then be estimated as 
\begin{equation}
\label{eq:deltafeta}
\Delta f \sim \hat{\Delta f} \equiv \text{Im}\left[\text{S}_{21}(\Delta f) \times \eta\right]
\end{equation}
\noindent where S$_{21}(\Delta f)$ is the complex transmission at the 
shifted resonance frequency. 
Throughout this text we denote estimates with a hat.
This estimate is equivalent to projecting the rotated and scaled S$_{21}$ 
response at the probe tone frequency onto a single axis. 
A sketch of this calibration scheme is given in Fig.~\ref{fig:etacal}. 

\begin{figure}
\includegraphics[width=0.95\columnwidth]{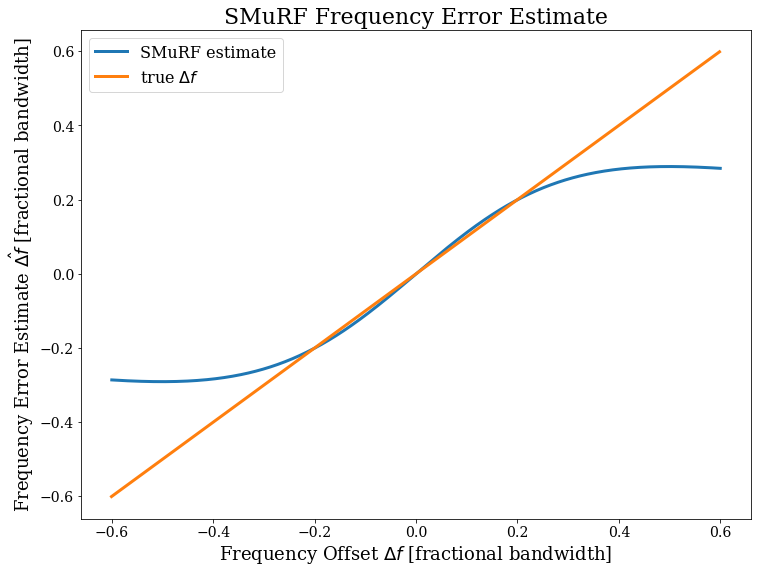}
\caption{\label{fig:deltafeta} Estimated frequency error with the $\eta$ 
multiplication (blue) versus real frequency error for a simulated resonator 
(orange). 
The regime in which the estimated frequency error $\hat{\Delta f}$ (blue) 
matches the real frequency error is in a narrow band about the resonator 
frequency ($\Delta f = 0$). 
SMuRF tracks resonators with tones in this narrow bandwidth while flux ramping 
using an adaptive filter as described in Section~\ref{sec:trackingdemod}.}
\end{figure}

The estimate holds for small changes in frequency and in the limit that shifting 
the resonance frequency away from the tone is the inverse of shifting the tone, ie
\begin{equation}
\label{eq:dS21df}
\frac{\partial\text{S}_{21}\left(f - f_\text{res}(\phi)\right)}{\partial f} = -\frac{\partial\text{S}_{21}\left(f - f_\text{res}(\phi)\right)}{\partial f_\text{res}(\phi)}
\end{equation}
\noindent where $f$ is the probe tone frequency and $f_r$ is the resonance frequency, 
which depends on the flux $\phi$. 
This assumption is valid for most resonator-based readout systems in the small 
signal limit, i.e. where the frequency shift $\Delta f$ is small compared with 
the resonance bandwidth. 
Figure~\ref{fig:deltafeta} shows the estimated frequency error $\hat{\Delta f}$ 
using the $\eta$ calibration versus the true frequency error $\Delta f$. 
In particular, we note that the frequency error estimate holds only for a small 
region within the resonance bandwidth.

The $\eta$ estimate must be performed once per resonance, typically at the first 
cooldown for a set of resonators. 
The calibration is largely stable between data-taking sessions and fridge cycles 
for a given set of devices, and saved for reuse between device cooldowns. 
Subsequent to $\eta$ calibration, the SMuRF is capable of channelizing tones 
and returning the output tone frequency and frequency error, interpreted as the 
resonator frequency offset from the probe tone, for each resonator simultaneously.

\subsection{\label{sec:trackingdemod}Tone Tracking and Flux Ramp Demodulation}
We implement a feedback loop to minimize the SMuRF estimate of
the frequency error magnitude and continuously update the probe tone 
to be positioned at the estimated resonance frequency. 

When using the microwave SQUID multiplexer, the detector signal is 
reconstructed via phase shifts in the flux ramp response, which are 
linear in detector signal for detector signals much slower than the 
flux ramp rate. 
For recent generations of NIST CMB-style \umux resonators with bandwidth 
of about 100~kHz, we typically modulate the SQUIDs with flux ramp sawtooth 
rates of several kHz and 3 to 6 flux quanta swept per sawtooth period.\cite{dober21}

Rather than design a feedback loop to directly track arbitrary signals
at 10s of kHz or faster, which is challenging due to latency from the
filter bank, we rely on the periodicity of the flux ramp response. 
That is, we express the flux-ramp modulated frequency response of the
resonator as in Equation~\ref{eq:freqvst}, reproduced here for
convenience:
\begin{equation*}
\Delta f(t) = B\left(\frac{\lambda\cos(\omega_ct+\theta(t))}{1+\lambda\cos(\omega_ct+\theta(t))}\right).
\end{equation*}
The carrier frequency $f_c = \omega_c/2\pi$ is the product of the flux ramp 
sawtooth reset rate and the number of flux quanta swept in a single flux ramp 
period.
For NIST CMB-style \umux resonators, the design values are a resonator peak
to peak swing of about 100~kHz and $\lambda\sim 1/3$, though this
discussion applies without loss of generality to any sufficiently
periodic signal.\cite{dober21}
For detector signal $\theta(t)$ approximately constant relative to
$\omega_ct$, we treat this as an estimation problem for multiple
sinusoids in a truncated Fourier series, where we have $M$ sinusoids
of known frequencies.
These frequencies are harmonics of the carrier with known frequency $f_c$ 
but unknown in amplitude or phase.

Thus, the resonance frequency of a given channel as function of time is 
parameterized by
\begin{equation}
\label{eq:freqest1}
f[n] = \text{constant} + \sum_{i=1}^M A_i\sin(\omega_i T_s n + \theta_i) + w[n],
\end{equation}
\noindent where we include a constant offset term and a Gaussian measurement noise $w[n]$. 
Here, $n$ is the sample number and $T_s$ is the time per discrete sample, which is 
$T_s = 1/(2.4~\mathrm{MHz})$ as defined by the filter bank per-channel bandwidth. 
Expanding the sine terms, we can rewrite the frequency as a sum of sines and cosines:

\begin{equation}
f[n] = \text{constant} + \left(\sum_{i=1}^M a_i\sin\omega_i T_s n + b_i\cos\omega_i T_s n\right) + w[n] \label{eq:freqest2}
\end{equation}
\noindent{where}
\begin{equation}
a_i = A_i\cos\theta_i,\quad b_i = A_i\sin\theta_i.\label{eq:coeffs}
\end{equation}
\noindent In the limit of slowly varying $\theta(t)$, the $a_i$ and $b_i$ are 
approximately constant, and the overall frequency estimate is linear in the 
$a_i,b_i$, and constant offset. 
The change in phase of each harmonic is given by $\Delta\arctan(b_i/a_i) = i\times\Delta\theta_i$. 
As the detector signal modulates the phase of the entire SQUID response curve, we 
seek an estimate of the phase of the fundamental harmonic $\theta_1 = \arctan(b_1/a_1)$.
Due to FPGA resources, in the \umux system we take the number of
tracked harmonics $M$ to be 3.
In practice this provides a sufficient parameterization of well-behaved SQUID
curves on NIST CMB-style \umux devices.  

For $M$ harmonics of a fundamental frequency $\omega_1$, Eq.~\ref{eq:freqest2} may 
be rewritten as

\begin{equation}
\label{eq:freqest3}
\vec{f} = \textbf{H}\vec{\alpha} + \vec{w}
\end{equation}

\noindent where $\vec{f}$ and $\vec{w}$ are $M\times 1$ column vectors of resonance 
frequency values and measurement noise for $N$ discrete samples per flux ramp frame.
We define coefficient vector $\vec{\alpha}$ as a $(2M+1)\times 1$ column vector of 
sine and cosine coefficients $a_1,b_1,\ldots, a_N,b_N,\text{constant}$ for $M$ 
harmonics and a constant offset term. 
The harmonic sample matrix $\textbf{H}$ is an $N\times (2M+1)$ matrix given by
\begin{widetext}
\begin{equation}
\textbf{H} = \left(\begin{array}{cccccc}
\sin_1[0]&\cos_1[0]&\sin_2[0]&\cos_2[0]&\cdots&1\\
\sin_1[1]&\cos_1[1]&\sin_2[1]&\cos_2[1]&\cdots&1\\
\vdots&\vdots&\vdots&\vdots&\ddots&\vdots\\
\sin_1[N-1]&\cos_1[N-1]&\sin_2[N-1]&\cos_2[N-1]&\cdots&1
\end{array}\right),\label{eq:defH}
\end{equation}
\end{widetext}
\noindent where

\begin{equation}
\label{eq:defsin}
\sin_m[n] \equiv \sin\left(\omega_m T_s n\right),\quad \omega_m = m\times\omega_1
\end{equation}

\noindent and similarly for $\cos_m[n]$.
Our problem thus reduces to estimating the coefficients vector $\vec{\alpha}$.

The current SMuRF tracking and demodulation loop implements this
estimation using a recursive algorithm.
The fundamental tracking frequency $f_1 = \omega_1/2\pi$ is assumed
known and must be measured using software tools (\S~\ref{sec:pysmurf})
or input by the user.
At each discrete timestep $n$, a probe tone centered on a resonance 
has frequency predicted as
\begin{equation}
\label{eq:freqpred}
\hat{f}[n] = a_1[n]\sin(\omega_1[n]) + b_1[n]\cos(\omega_1[n]) + \cdots + C
\end{equation}
\noindent which may be thought of equivalently as the dot product of the 
$n$th row of $\textbf{H}$ with $\vec{\alpha}$. 
Simplifying notation, we can write the $n$th row of $\textbf{H}$ as $\vec{h}[n]$, given by
\begin{equation}
\label{eq:defh}
\vec{h}[n] = \left(\begin{array}{cccccc}\sin_1[n] & \cos_1[n] & \cdots & \sin_N[n] & \cos_N[n] & 1\end{array}\right).
\end{equation}
Eq.~\ref{eq:freqpred} is then written more compactly as
\begin{equation}
\label{eq:freqpred2}
\hat{f}[n] = \vec{h}[n]\cdot\vec{\alpha}[n].
\end{equation}

We measure the frequency error as described in \S~\ref{sec:eta} following 
Eq.~\ref{eq:deltafeta} as 
\begin{equation}
\label{eq:freqerror}
\hat{\Delta f}[n] \equiv f[n] - \hat{f}[n] \approx \text{Im}\left(S_{21}\left(\hat{f}[n] - f_\text{res}[n]\right) \times\eta\right)
\end{equation}
Importantly, we measure only $\hat{\Delta f}[n]$ and do not have access to 
the true $f_\text{res}[n]$. 

Based on this readback, we update the prediction via a feedback loop such that
\begin{equation}
\label{eq:updatepred}
\vec{\alpha}[n+1] = \vec{\alpha}[n] + \mu\hat{\Delta f}[n](\vec{h}[n])^T
\end{equation}
where $\mu$ is a user-tuned gain parameter. 
This algorithm is equivalent to the stochastic gradient descent method, where 
we are attempting to minimize $\hat{\Delta f}[n]$.\cite{sgd}
Note that $\hat{f}[n]$ and $\hat{\Delta f}[n]$ are scalars corresponding to the 
probe tone frequency and estimated frequency error, respectively, of a single 
channel at discrete timestep $n$. 
Since $\omega_1$ is known, the flux ramp rate is no longer limited by
the bandwidth of the constant-only tracking loop, i.e. the $N=0$ case.

In this $N=0$ case of no flux ramp such as with KID systems or when 
slower tracking is desired, $\vec{\alpha}$ reduces to a
single constant frequency error $\alpha = \Delta f$.
We further see that $\vec{h}[n]$ is simply 1, and $\hat{\Delta f}[n]$ is 
the estimated frequency error, which is added to the previous constant 
offset term $\alpha$ with some gain $\mu$.
In this limit, a model with only the constant offset term is 
equivalent to having only proportional control in the tracking loop. 
The bandwidth of this operation alone is about several kHz, setting the
limit for tracking bandwidth in non-flux ramped systems.
Planned modifications for KID systems would use the resources freed up by 
no longer using the higher harmonic tracking to add a proportional gain 
term or a lead/lag compensator to increase the loop bandwidth, allowing 
for higher bandwidth applications such as calorimetry. 

Since this tracking loop estimates coefficients of the harmonic
content, the constant term encodes frequency shifts not directly
sampled by the harmonics vector $\vec{h}[n]$.
In the case of poorly selected tracking frequencies $\omega_i$, this
effectively aliases against the true modulation frequency and
misconstructs the phase.
The constant term $C$ absorbs slow shifts in the resonator frequency
or the amplitude of transmitted quadrature voltage through the
resonator, potentially suppressing sources of noise including but
not limited to amplifier gain fluctuations and microphonic pickup. 

The user-defined parameters to the tracking algorithm are the fundamental 
harmonic frequency $\omega_1$ and the gain $\mu$. 
As is typical for generic tracking feedback systems, the gain must be large 
enough to guarantee convergence without causing stability issues. 
In practice, there exists a broad range of gain values for which the
feedback loop is stable, with the gain determining the feedback loop
bandwidth.

The fundamental tracking frequency $f_1$ is independent of the
resonator bandwidth, and resonators centered in bins of the polyphase 
filter bank can in principle be tracked as fast as the per-channel 
Nyquist frequency, which for the SMuRF CMB firmware implementation is 
1.2~MHz.
Resonator tracking at rates faster than the resonator bandwidth may be
desirable for several applications.
The tracking loop updates for each channel at the per-channel
digitization rate of 2.4~MHz; however, the stochastic gradient descent 
method is only expected to converge on the mean phase over the 
flux ramp period.~\cite{mandic15} 
Therefore, while the tracked frequency of each resonator is sampled at
2.4~MHz, detector information is only sampled at the flux ramp rate.
Because the tracking feedback loop acts as a low pass filter, 
its bandwidth must be limited to at most half the flux
ramp rate through the choice of tracking gain $\mu$ or pre-filtering 
to avoid aliasing detector noise into the signal band.
Combined with the fact that increasing flux ramp rate modulates the 
detector signal band to a region of lower TLS contributions, higher flux 
ramp rates are generically preferred to reduce readout noise.~\cite{ltd21}

To avoid losing lock on the feedback loop during the flux ramp sawtooth 
resets, which cause a transient in the SQUID that are difficult to track, 
the tracking algorithm supports a user-tuned window within each 
flux ramp frame to be ``blanked off'', or removed from the feedback loop. 
During this period, the coefficients $\alpha$ are held constant at the 
last updated value. 
The size of this window is typically tuned to cover the sawtooth 
reset period and subsequent transient response, and to maintain an 
integer number of flux quanta being tracked in the active loop. 
Blanking introduces an aliasing penalty due to the lower duty cycle,
which is mitigated by keeping the blanking window size as short as
possible.

\subsection{\label{sec:misc}Other Miscellaneous Firmware Functions}

In addition to tone synthesis and readback, channelization, frequency error estimation, 
and tracking, the firmware handles miscellaneous functions such as the data
streaming interface, timing synchronization between SMuRF and external
systems, and control and configuration of the AMCs, RTM, and cryostat
card.

In the carrier FPGA firmware, the streaming data framer sums the
estimated coefficients of the first tracked harmonic $a_1,b_1$ as
discussed in \S~\ref{sec:trackingdemod} for each flux ramp frame and
writes them to a dual port RAM at the start of each frame. 
This effectively averages over the flux ramp frame. 
From the dual-port RAM of each baseband processor, the per-channel 
coefficients and channel ID information are combined into a full 4096 channel RAM. 
The per-channel phase is computed serially as $\theta_1 =
\arctan(b_1/a_1)$ with a coordinate rotation digital computer (CORDIC) 
algorithm, packaged with header information, and passed to the
software streaming interface.
The header includes information about the hardware configuration
including the RTM DAC settings including detector bias settings, flux
ramp status, timing counters and synchronization bits, data rate, and
optional user words.
Additional coefficients from the higher harmonics of the tracking loop
could in principle be added to the data stream for higher signal to noise 
performance, although in practice have not yet been found necessary and are 
not currently implemented.

The arctan computation is clocked at 156.25~MHz.
Since all 3328 channels must pass through a single CORDIC for the phase calculation, 
this clock thus limits the per-channel streaming data rate to 156.25~MHz / 3328 
channels $\sim$ 50~kHz. 
Since the flux ramp reset rate in typical operation has not exceeded 
$\sim 30$~kHz, this is acceptable for current uses but could be upgraded 
to faster clock rates if necessary.
Faster data rates are available for writing directly to disk for diagnostic and 
debugging purposes.
These may be sampled up to the per-channel digitization rate of 2.4~MHz per 
channel, or the decimated 614.4~MHz ADC speed for the full band without 
channelization. 
Applications with fewer channels, such as high bandwidth pulse detection 
applications, would also have higher data streaming rates without the need for 
an upgraded clock. 

The flux ramp sawtooth reset is triggered through one of three possible modes: 
(1) internally, (2) through a trigger from a common timing system, or (3) from an 
external TTL trigger on a LEMO port on the RTM.
The internal timing mode uses a 32-bit programmable count limited
counter in the FPGA, clocked at the complex baseband processing rate of 307.2~MHz.
The trigger and pulse width are passed to the digital signal
processing core for flux ramp demodulation and to the CPLD, which has
a 32-bit counter to receive the external reset.
The upper 20 bits of the CPLD counter are selected to output to the
flux ramp DAC on the RTM, which allows for better flux ramp amplitude
control.
For large installations in which the timing information is important,
the external timing option allows for higher precision and
synchronization.
The common timing system was originally developed for accelerator systems and
designed for synchronizing large numbers of devices to sub-nanosecond
precision.\cite{frisch16}
In this case, a separate timing card distinct from the SMuRF 
carrier card generates a timing data stream at 2.4576~GS/s and 
broadcasts timing words over a serial fiber.
The timing words are subsequently decoded by each individual SMuRF carrier card.
These words may be derived from an external timing system to allow for
synchronization between SMuRF and non-SMuRF timing systems within a
given experiment.

While the bulk of the firmware is located on the carrier card, the PIC
microcontroller on the cryostat card and the CPLD on the RTM
additionally run firmware.
The RTM CPLD is nominally clocked at 51.2MHz to accommodate the maximum frequency 
of the flux ramp DAC of 50MHz, though the flux ramp DAC is typically operated well 
below this frequency. 
An arbitrary waveform generation module containing a look-up table is capable of 
writing pairs of bias DAC values sequentially, which allows for up to $\sim 1$kHz 
updates to the DAC output voltages.
Due to firmware resource limitations, the CPLD interface is write-only, so to 
maintain a record of operations we cache written values in FPGA shadow memory (BRAM).

The large-scale PCIe streaming interface splits incoming data from the QSFP links 
to separate ports handling firmware register access and streaming data, respectively. 
Up to 6 carriers, typically run together in a single 7-slot ATCA crate with a 
network switch occupying the final slot. 
Each carrier's register access and data streaming ports are routed to the CPU via 
direct memory access (DMA) lanes. 
The register access port is low-latency for fast commanding and access. 
Every pair of carriers receives a dedicated 4~GB error correction code (ECC) DDR4 
block, allowing for 2~GB of deterministic buffering per carrier before the CPU RAM 
to avoid dropping packets due to CPU back pressures on the DMA path. 
To avoid software data operations from back pressuring the register access, 
the DMA lanes for register access and data streaming are kept separate with independent 
PCIe DMA descriptors. 

\subsection{\label{sec:resources}Firmware Resource Consumption}

\begin{figure}
\includegraphics[width=0.99\columnwidth]{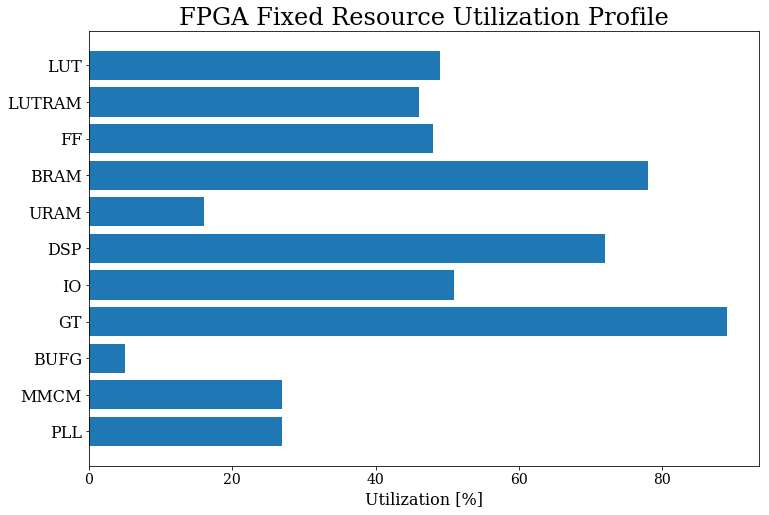}
\caption{\label{fig:fpgaresources} Total consumption profile of fixed
  resources for the FPGA, a Xilinx Kintex UltraScale+ XCKU15P. 
  The terms are as follows: LUT = look-up tables, LUTRAM =
  look-up table RAM, FF = flip-flops, BRAM = block memory, URAM =
  larger block memory, DSP = digital signal processing blocks, IO =
  input/output interfacing, GT = gigabit transceivers, BUFG = global
  clocking, MMCM = mixed-mode clocking manager, PLL = phase-locked
  loops for clocking.}
\end{figure}

A challenge for SMuRF was fitting the firmware within the
resource profile of the FPGA.
The total FPGA resource consumption profile is given in
Figure~\ref{fig:fpgaresources}.
We see that the most intensive elements for the SMuRF involve memory
and signal processing.
The high FPGA firmware resource usage and high 614.4~MHz clock rate is
near the device limits and require substantial firmware optimization.
We confirm that the requisite power draw and thermal environment can
be maintained within device limits for this configuration even in
high-altitude environments in \S~\ref{sec:perfaltitude}.

\section{\label{sec:soft}Software}
\begin{figure}
\includegraphics[width=0.5\columnwidth]{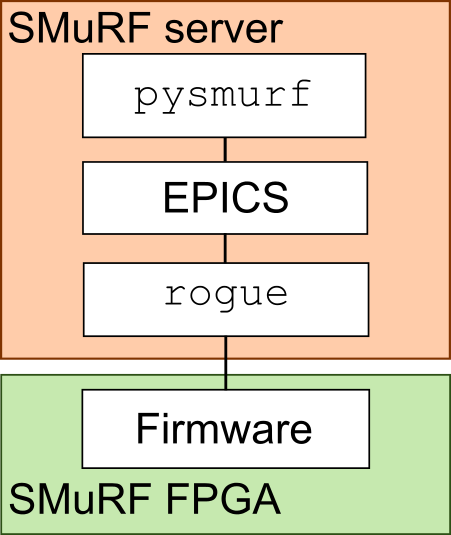}
\caption{\label{fig:software_block} A block diagram of the software components. 
A server (Dell R440) interfaces with the FPGA via Rogue via the EPICS framework. 
The \texttt{pysmurf} control software provides the end-user interface with the system.}
\end{figure}

A block diagram of the SMuRF software interface is given in Fig.~\ref{fig:software_block}. 
The FPGA is interfaced with a server via Rogue\footnote{\url{https://github.com/slaclab/rogue}}, 
a Python-based hardware abstraction and data-streaming system. 
Rogue registers are accessed via the Experimental Physics and Industrial Control 
System (EPICS)\footnote{\url{https://docs.epics-controls.org/en/latest/guides/EPICS\_Intro.html}}, 
a control system architecture which acts as a communication protocol between the firmware 
registers and the end user's desired interface. 
For SMuRF end users, the EPICS layer is further wrapped in the open source 
\texttt{pysmurf} control software, which provides higher level processing and 
analysis tools.\footnote{\url{https://github.com/slaclab/pysmurf}}
Each of these layers is discussed in turn below. 

To guarantee a consistent software environment across multiple deployed systems, 
the software suite is version controlled with \texttt{git}\footnote{\url{https://git-scm.com/}} 
and deployed to user institutions via Dockers\footnote{\url{https://www.docker.com/}}.
Furthermore, each hardware component has a serial number programmed in that may be 
remotely queried to facilitate hardware tracking across multiple deployments.

\subsection{\label{sec:rogue}Rogue}
The Rogue system serves as the communication layer between the FPGA firmware and 
the server software. 
It provides data streaming and hardware abstraction in a clean, modularized format 
that enables systems to be brought up from the prototyping phase to the system 
integration and deployment stage.
Firmware registers are accessed via a hierarchical tree structure that encodes and 
optionally commands the configuration and status of each register. 
A memory API and data stream API separately handle FPGA memory access and data 
processing, respectively. 
The data stream may be passed directly to downstream DAQ programs such as telescope 
control suites, as described in \S~\ref{sec:telescopescontrol}.
Metadata streams of any Rogue process variable can also be streamed alongside the 
data at the data rate, allowing for real time recording of important calibration 
and state information. 

\subsection{\label{sec:epics}EPICS}

The Rogue firmware registers are accessed via Experimental Physics and Industrial 
Control System (EPICS) process variables.\footnote{\url{https://epics-controls.org}}
The EPICS interface provides a control system architecture with a distributed 
real-time database of register values and a client/server model with efficient 
communication protocol for passing data. 
Existing infrastructure for EPICS is built on decades of experience in designing 
and implementing large-scale control systems for scientific instruments such as 
accelerator systems, including pre-built software tools. 
It is currently actively developed and maintained by an international collaboration 
of users across several large scientific facilities. 
A channel access protocol connects the EPICS server, which wraps the Rogue registers, 
to the client-side process variables that may be queried, set, or monitored by any 
computer on the same client network. 
EPICS process variable interaction may be conducted in essentially any modern 
interface, including but not limited to C/C++, Java, perl, python, Tcl/Tk, sh, 
and Matlab. 

\subsection{\label{sec:pysmurf}\texttt{pysmurf} Control Software}
Since python is an open source language used extensively for astronomical 
applications and laboratory experiment settings, our primary user software 
suite is built in Python3. 
The \texttt{pysmurf} control software serves as the interface for general users 
interacting with the SMuRF system.
The software provides the following functionality:
\begin{itemize}
    \item Wraps EPICS process variable control functions and provides data handling 
    and low level commands, such as turning tones on and off, setting DAC voltages 
    for detector and amplifier bias, and acquiring time-ordered data.
    \item Provides utility functions for finding and characterizing resonators, 
    setting up tone tracking, and acquiring flux ramp demodulated data.
    \item Stores specific hardware configuration information such as in-line wiring 
    resistance, band frequencies, SQUID couplings, and shunt resistances in a 
    human-readable configuration table that may be edited or referenced in later analysis.
    \item Offers higher level control and analysis functions, including but not limited 
    to noise analysis, TES detector characterization, and system optimization.
\end{itemize}
The repository is \texttt{git} controlled and actively managed at SLAC to ensure 
continuity and cross-system compatibility as the software advances. 

\subsection{\label{sec:telescopescontrol}Integration with Telescope Control Suites}

The \texttt{pysmurf} software suite and Rogue streaming interface can
be integrated with other experimental DAQ/control suites, aligning
detector data with information such as telescope pointing,
housekeeping, and observatory state.
It is possible to either encode syncwords in the frame headers or
release frames based on sync signals for frame alignment.
These syncwords, typically received from an experiment common timing 
synchronization system, allow for frames to be aligned offline, typically by 
a downstream frame packager or in post-processing. 
This integration has been successfully demonstrated with the
\texttt{generic control program} (gcp) observatory control suite in a
test deployment with the BICEP/Keck program, and has been built for
the Observatory Control System (OCS) in anticipation of deployment for
the Simons Observatory.\cite{story12,cukierman19,koopman20}
Critical setup, calibration, and control functions may be received by
SMuRF from a central telescope DAQ.
Optionally downsampled data is digitally low-passed with a multi-pole
Butterworth filter with user-defined rolloff frequency.
Useful registers may additionally be encoded in frame headers,
allowing for streaming of information such as detector bias levels and
observatory timing synchronization.

\subsection{\label{sec:babysmurf}Simulation Suite}
To model the impact of the SMuRF firmware algorithms, particularly in 
frequency error estimation and flux ramp demodulation as described in 
\S~\ref{sec:eta} and \ref{sec:trackingdemod}, these algorithms are implemented 
in a publicly available Python simulation suite.\footnote{\url{https://github.com/cyndiayu/babysmurf}}
The simulation suite takes as input real or simulated resonator and SQUID characteristics, 
detector-referred input signals, and readout parameters such as flux ramp rate and 
tracking loop gain, and outputs the flux ramp demodulated detector phase as would be 
output by SMuRF. 
The details of this code and example studies of the impact of various user-selected 
parameters on final detector-referred demodulated data are presented elsewhere.\cite{babysmurf}
Integration with the \texttt{pysmurf} user control software suite is planned for a future release. 

\section{\label{sec:perf}Performance}

SMuRF systems have been successfully integrated and used for both
laboratory and field measurements of CMB detectors, demonstrating
achievement of end-to-end system performance meeting the goals of CMB
experiments.\cite{cukierman19,dober21,mccarrick21}
In this section we explicitly address the electronics requirements as
outlined by \S~\ref{sec:req}.

\subsection{\label{sec:perfdyn}RF Dynamic Range and Linearity}

\begin{figure*}
  \includegraphics[width=0.99\textwidth]{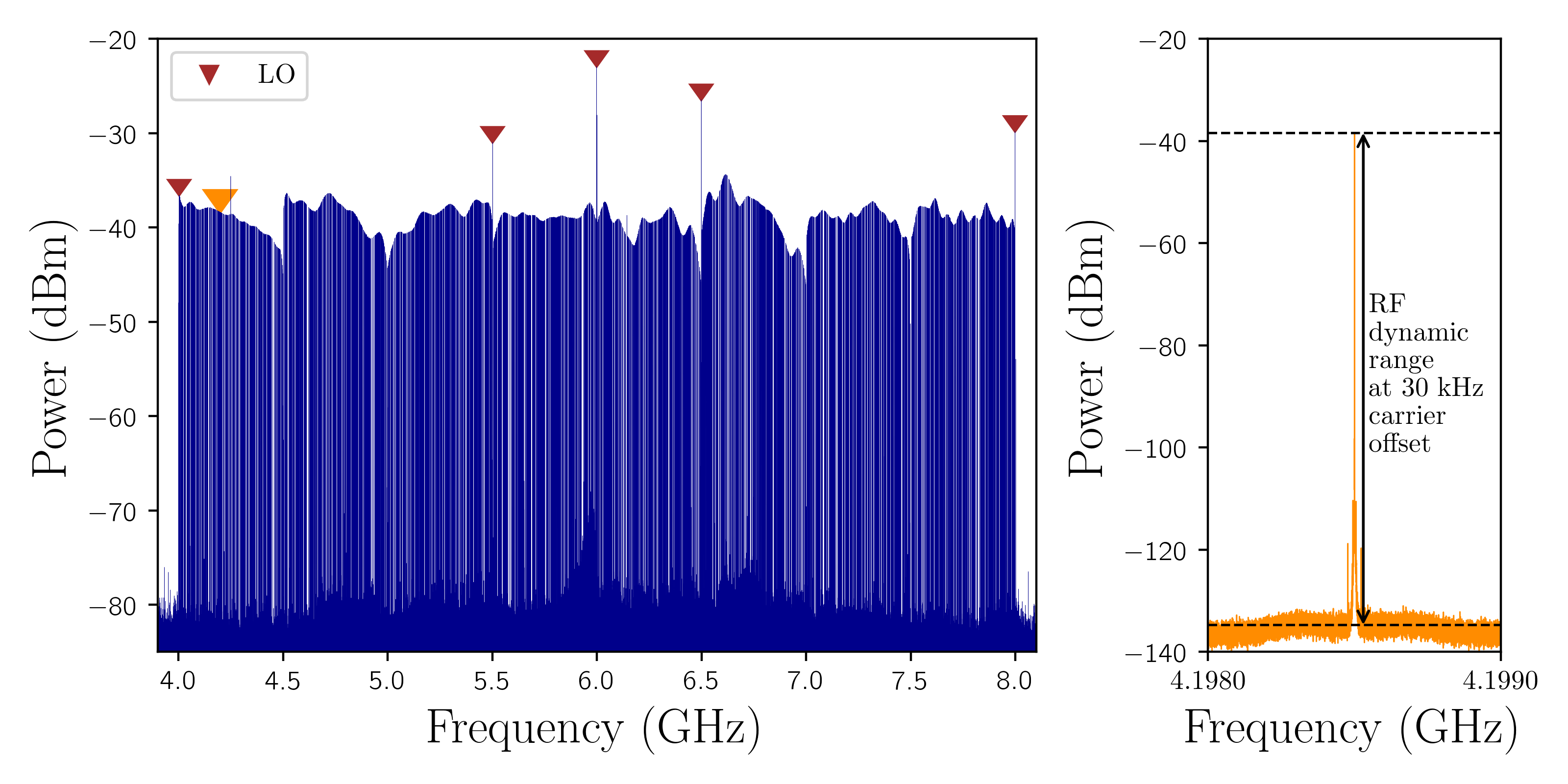}
  \caption{\label{fig:dynrange} [Left] 2000 fixed tones generated with SMuRF 
  over 4--8~GHz (with 250 tones in each 500~MHz band) measured directly using 
  a signal analyzer.  
  LOs which fall within the frequency range of the plot are labeled with smaller brown markers.  
  The average tone power in each 500~MHz band has been leveled using the programmable output 
  attenuators.
  [Right] Zoom in of a single tone while playing 2000 fixed tones, indicated by a larger 
  orange marker on the lefthand plot.  The arrow indicates the measured RF dynamic range 
  at a $30$~kHz carrier offset.  The signal analyzer trace has been median filtered 
  (while preserving the maximum) to suppress narrow-band spurs.
  The dynamic range at $30$~kHz carrier offset exceeds the SMuRF requirement of 
  $>100$ dBc/Hz after accounting for the signal analyzer noise floor (which is only 
  $\sim$6~dB lower) and noise which cancels when the RF signals are downmixed with 
  their common LO.}
\end{figure*}

As outlined in \S~\ref{subsec:noise}, the RF dynamic range of the
electronics in dBc/Hz, or the level of the total RF noise floor, is
critical to the readout white noise performance.
The SMuRF was designed to achieve better than 100~dBc/Hz at 30kHz
carrier offset for 2000 tones.
At smaller carrier offsets, the dynamic range is expected to degrade due to 
the noise profile of the LO used to upmix the synthesized tones. 
The dynamic range of the SMuRF system is considered through the full
tone synthesis and analysis chain, which includes the ADCs, DACs, 
and the polyphase filter banks.
In Figure~\ref{fig:dynrange} we show 2000 tones generated over
4--8~GHz by a SMuRF system and the dynamic range of one of those tones
measured directly using a signal analyzer connected to the SMuRF RF
output with no additional amplification or attenuation beyond a short
length of coaxial cable.
Note that measurements do not include the effects of the downmix,
ADCs, or analysis polyphase filter bank which combined can both
improve and degrade the achieved end-to-end RF dynamic range.

\begin{figure*}
\includegraphics[width=0.99\textwidth]{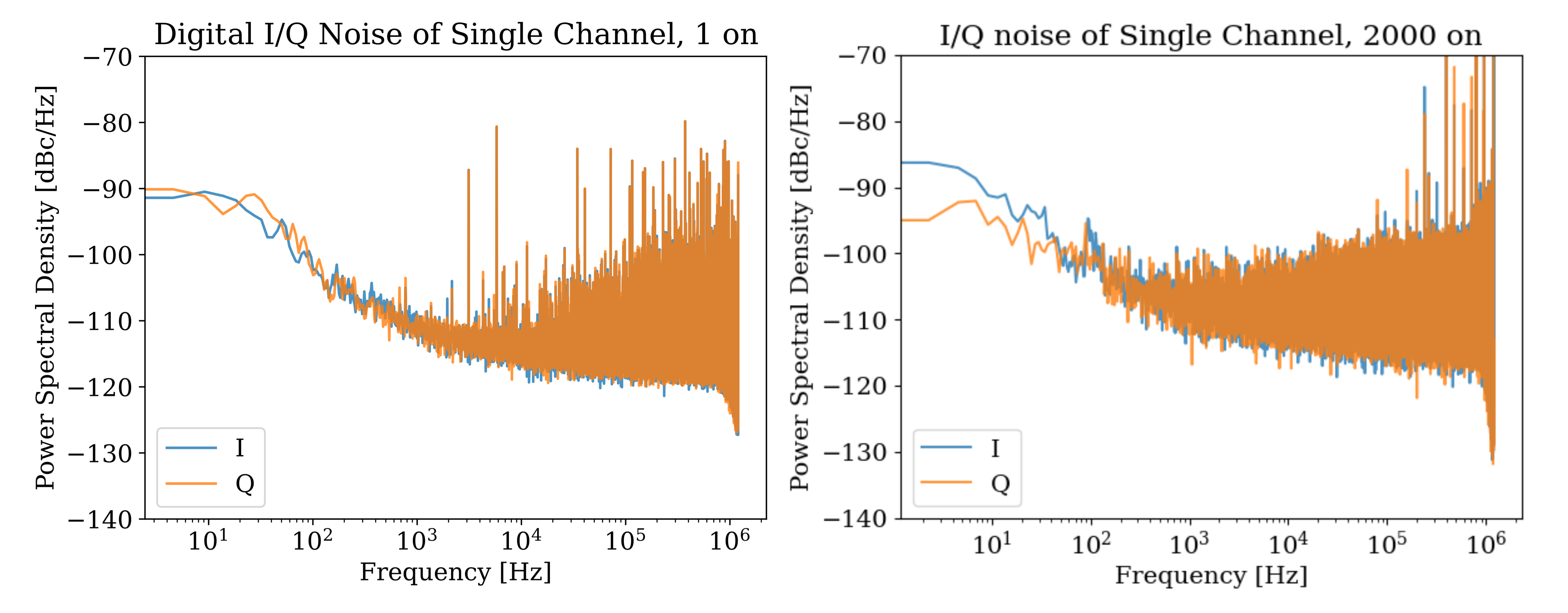}
\caption{\label{fig:psd} [Left] The noise spectral density of a single
  channel looped back directly from RF out to RF in with only that
  channel on. The traces represent orthogonal data components I and Q,
  which should be identical given there is no resonator in the
  measurement. We expect the $1/f$ knee to be at several hundred Hz due to
  the LO generation and RF mixers. The white noise level is below
  -100~dBc/Hz, as expected, with a roll-off at 1.2~MHz defined by the
  Nyquist rate of the polyphase filter bank. Narrow-band spikes are
  due to a combination of hardware and firmware DSP effects, but do
  not contribute substantially to the total noise power. 
  [Right] The same measurement as the left, but with 2000 tones turned on 
  from 4-8~GHz. The minimal degradation in noise performance demonstrates 
  the excellent linearity of the full SMuRF system. 
  }
\end{figure*}

To check the impact of the entire digital processing chain,
we acquire a 0.5 second timestream of the digitized I and Q orthogonal 
data components from a single channel with a low-band AMC connected in loopback, 
i.e. with a short coaxial cable connecting its RF input to its
RF output. 
Since the noise performance of a single channel is limited by firmware rather 
than hardware choices, one AMC suffices to demonstrate performance without 
the need for a second AMC. 
The tone power is set to -33~dBm output from the DAC, with an
additional 6~dB of programmable RF attenuation added to each 
the RF input and output on the AMC, as is typically used for NIST 
CMB-style resonators. 
The digital I/Q data are acquired without additional processing, corresponding 
to the state of the data at the output of the analysis filter bank as described 
in \S~\ref{sec:pfb}. 
We take the power spectral density of the time-ordered data and convert to units of dBc/Hz 
via normalizing by the magnitude of the time-ordered data stream. 
This is given in the left panel of Figure~\ref{fig:psd}. 

\begin{figure*}
\includegraphics[width=0.99\textwidth]{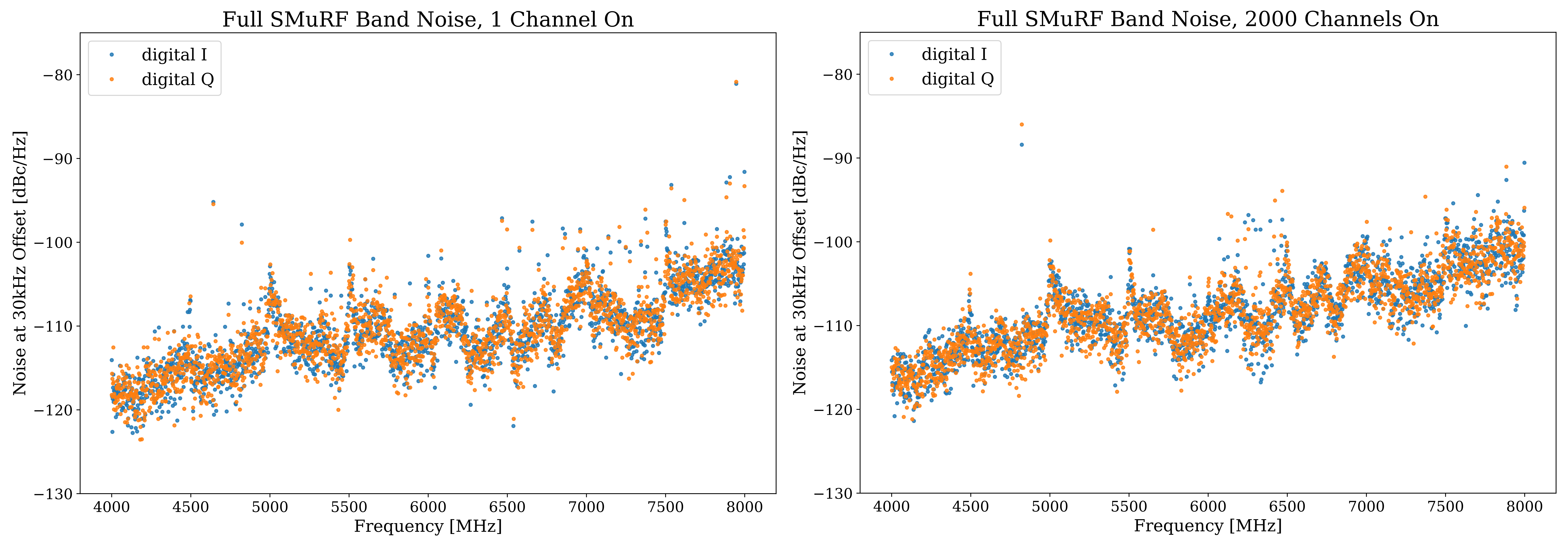}
\caption{\label{fig:profile} [Left] The noise at 30~kHz carrier offset of a 
single channel in loopback across the entire 4-8~GHz frequency range. 
We see that the overall level remains below -100~dBc/Hz aside from a handful of 
outliers. 
The edges of each 500~MHz band are visible due to the slight roll-off in the 
edges of the passband of the cavity filters. 
A small overall tilt is expected due to the degradation of RF performance at 
higher frequencies for these broadband components. 
[Right] The same measurement taken while 2000 tones are turned on across 4-8~GHz, 
with 250 randomly chosen frequencies chosen in each 500~MHz band. 
We see that the overall level and shape are nearly identical to that of the 
single-channel case. }
\end{figure*}

We note that the $1/f$ performance of a single channel is due to $1/f$ noise in 
the hardware components, particularly in the LO generation and RF mixers, which 
have an $f_\text{knee}\sim 100$~Hz. 
However, for \umux this performance is irrelevant due to the flux ramp modulation 
scheme: we focus instead on the noise at the carrier offset, where the sidebands 
due to flux ramp modulation are sampled. 
Thus, we repeat the measurement shown in Figure~\ref{fig:psd} while sweeping the 
generated tone across the entire frequency band and extract the noise level at 
30~kHz offset, a typical flux ramp $\Phi_0$ rate. 
This is given in the left panel of Figure~\ref{fig:profile}.
For applications requiring better $1/f$ stability in RF performance, upgraded hardware 
components have been identified and may be swapped in easily. 

We see that the noise floor is well below the requirement of -100dBc/Hz across 
the entire frequency range. 
The noise rises slightly at the band edges, as is expected given the finite roll-off 
of the cavity filters on the AMC RF daughter cards.

A key design consideration for SMuRF is its ability to read out a
large number of channels for cryogenic multiplexing applications.
We trade off single channel noise performance for linearity across a large number 
of channels.
We therefore require that the noise performance achieved in
one channel is scalable to 2000 channels, corresponding to
the expected multiplexing density for NIST CMB-style \umux resonators. 
In the right panel Figure~\ref{fig:psd}, we show the noise spectral density of a
single channel in loopback while 2000 channels at -33~dBm each are
turned on from 4-8~GHz across two AMCs, which are chained in 
a configuration consistent with operating a full 4-8~GHz resonator 
set on a single coaxial line within the cryostat.
This corresponds to the maximum expected channel density of a full 
operating SMuRF system, limited by the density of cryogenic channels.  
We see that the noise performance suffers minimal penalty as compared
to the single channel case shown in the left side of the figure. 

\begin{figure}
\includegraphics[width=0.99\columnwidth]{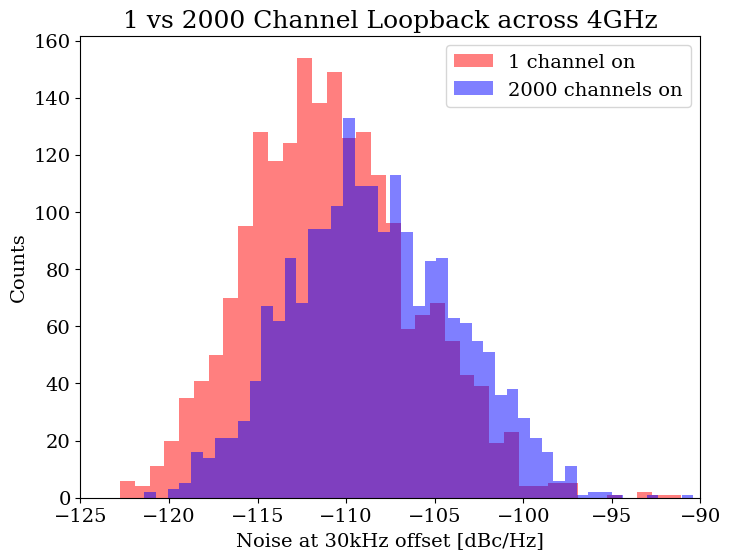}
\caption{\label{fig:profilehist} Noise at 30~kHz offset of each channel in loopback 
with 1 or 2000 channels turned on using identical attenuation and amplification settings. 
The 2000 channel data was taken with 1000 channels each from a low band and high band 
AMC connected serially in loopback. 
We see that there is a modest penalty from turning on 2000 channels across the 4~GHz 
bandwidth, but the overall distribution falls within the -100~dBc/Hz specification 
from the initial requirement. }
\end{figure}

Similarly, with 2000 channels turned on from 4-8~GHz on two AMCs, 
we take the noise on each channel serially 
and extract the noise level at 30~kHz offset.
This is shown in the right panel of Figure~\ref{fig:profile}, with the single
channel case for comparison given on the left panel. 
For more explicit comparison, in Figure ~\ref{fig:profilehist} 
we plot the histogram of the counts in each side of Figure~\ref{fig:profile}.

\subsection{\label{sec:perfwhiteandlowfreqnoise}White and Low Frequency Noise Performance}

Reading out newer, optimized (v3.2 or later) NIST CMB-style resonators not
connected to TESs in dark cryostats with basic magnetic shielding
(e.g. only a room-temperature mumetal can around the cold multiplexer
volume), \umux-coupled SMuRF systems routinely achieve white noise levels of  \Order(40~pA/$\sqrt{\mathrm{Hz}}$) TES current noise-equivalent with
an approximately $1/f$-like spectrum with $1/f$ knees of
\Order(10~mHz) while multiplexing \Order(1000) channels.\cite{henderson18,mccarrick21b}
For ground-based CMB receivers instrumented with 
\Order(10~m$\Omega$) normal resistance TESs this level of readout
noise is expected to increase the total per-detector noise-equivalent power, which 
is dominated by fluctuations in the observed photon intensity, by only an additional 1-2\%.
While a full analysis of the different noise contributions is deferred
to a future publication, typical measured white noise levels measured
at a \Order(30~kHz) $\Phi_0$ modulation rate are dominated
by resonator TLS noise, with a subdominant contribution from SMuRF’s
finite RF dynamic range.

Confirming the TES and flux ramp drives achieve their low $50$
pA/$\sqrt{\mathrm{Hz}}$ output current noise specification (see
Section~\ref{subsec:noise}) is difficult using conventional benchtop
instrumentation due to the required sensitivity.
From integrated system measurements, the estimated white noise
contributions from the TES and flux ramp drives are consistent with exceeding 
their output current noise specification.
At low frequencies, characterizing the noise performance of the TES
and flux ramp drives is further complicated by additional noise
sources including variations in passive component
temperatures, microphonics, and RF phase drift. 
A method for monitoring the overall phase drift of the RF line for slow changes 
to enable feedback on tuning parameters or offline data cleaning has 
further been developed.\cite{silvafeaver22}
Further discussion of the low-frequency system performance, with particular 
focus on the SMuRF TES and flux ramp drives as well as mitigation techniques 
integrated into the software and hardware, will be addressed in a future 
publication. 

\subsection{\label{sec:perftracking}Amplifier Linearity and Tone Tracking}

As detailed in \S~\ref{subsec:lin}, microwave resonator
systems are often limited by the linearity of both warm and cryogenic
RF components.
We see above in \S~\ref{sec:perfdyn} that the SMuRF RF components achieve 
the requisite linearity: the warm system in loopback suffers minimal 
degradation when 2000 channels are operated simultaneously. 
The linearity of the full RF chain is therefore typically limited by 
the achievable linearity of the cryogenic amplifier and other cold RF 
components. 
As described in \S~\ref{sec:trackingdemod}, SMuRF addresses this
linearity problem by implementing a closed loop feedback algorithm to maintain
the probe tone on the point of minimum amplitude transmission S$_{21}^{\mathrm{min}}$.

\begin{figure}
\includegraphics[width=0.99\columnwidth]{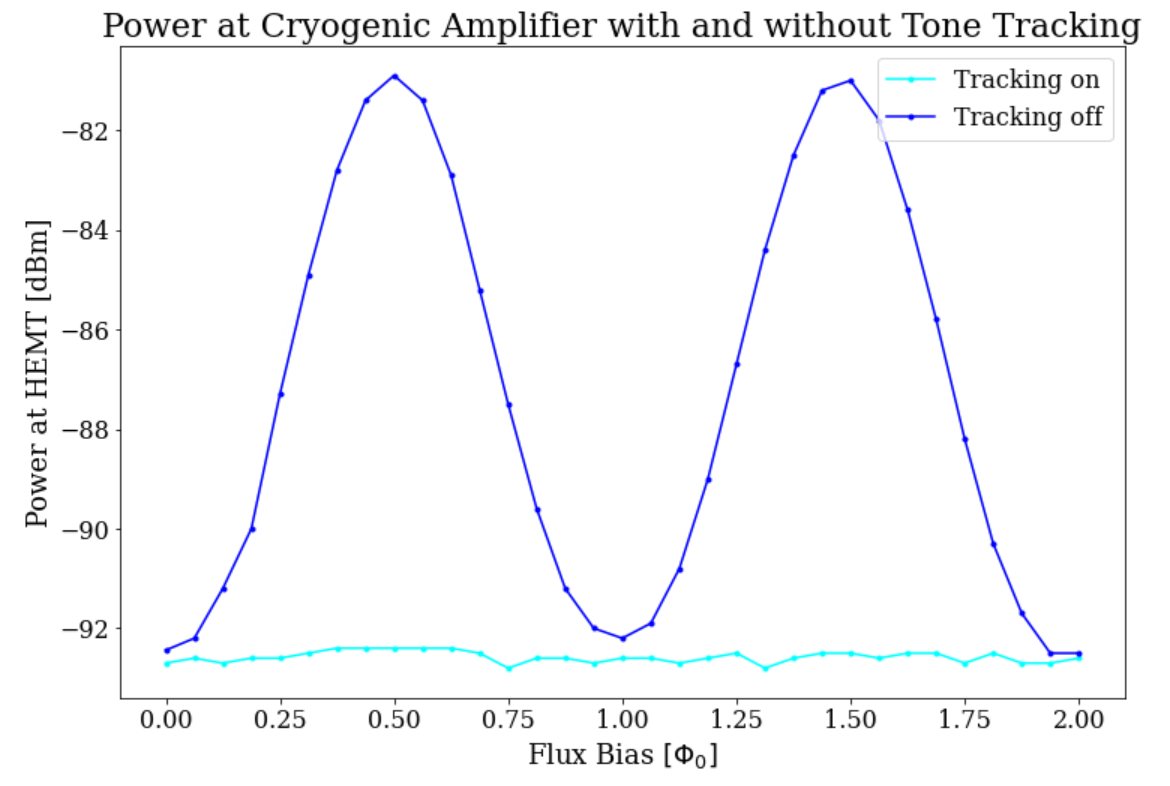}
\caption{\label{fig:hemt_power} The power incident on the first-stage cryogenic
  amplifier with and without tone tracking for a single tone tuned to a 
  NIST CMB-style \umux resonator. We see that the average power incident on the 
  amplifier is reduced by about 5~dB for a single tone due to tone tracking, 
  and by more than 10~dB for some bias points, significantly reducing the linearity 
  requirement for the cryogenic amplifier performance.
}
\end{figure}

We show that this tone tracking system succeeds in reducing the power
at the 4K cryogenic amplifier in a measured system in
Figure~\ref{fig:hemt_power}.
A resonator was calibrated using \texttt{pysmurf} software tools, with a directional coupler splitting the output of the cryostat between the SMuRF RF input on the strongly-coupled port and a spectrum analyzer on the weakly-coupled port.
The power at the input of the cryogenic LNA chain was computed by taking the integrated power in a
100~Hz window around the probe tone frequency as measured by the
spectrum analyzer and adjusting appropriately for the
amplifier gains and coupler loss.
When tracking is off, the probe tone is stationary and the total power
at the 4K LNA input varies by over 10~dBm.
After tracking is enabled, the probe tone moves with the resonance,
and the power incident on the cryogenic amplifier is minimized and
constant.
Since the system linearity scales exponentially with the power incident 
on the RF amplifiers, this allows for substantial relaxation on 
the requirements for the linearity of the hardware components. 

\begin{figure}
\includegraphics[width=0.99\columnwidth]{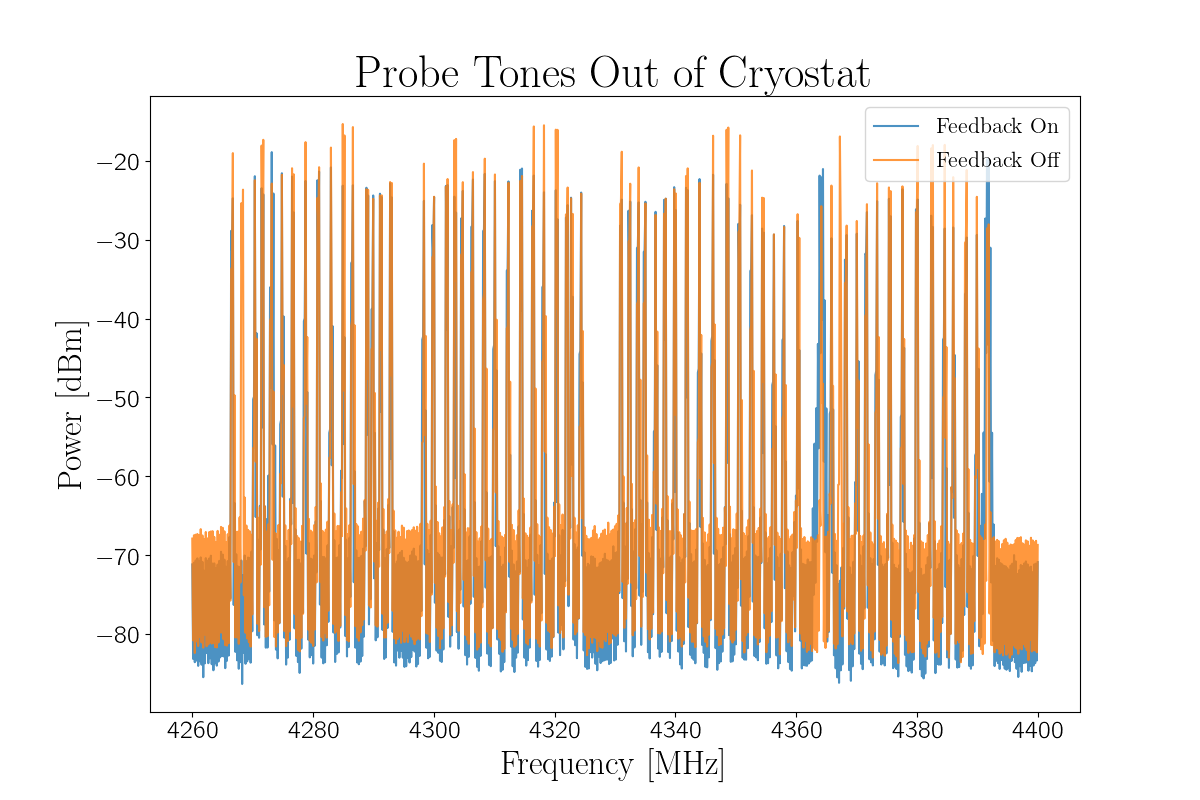}
\caption{\label{fig:tracking_onoff} Spectrum analyzer trace of the probe tone 
comb interacting with a single NIST CMB-style \umux resonator chip containing 65 
SQUID-coupled resonators. 
Probe tones approximately equivalent to -70~dBm per tone at the input of the 
resonators are calibrated to the resonators. The spectrum analyzer was placed 
at the equivalent of the RF input to SMuRF, after probe tones interact with the 
resonators and are amplified by 4K, 50K, and room temperature amplifiers. 
No additional flux is applied; thus, any change in resonance frequencies is due 
to noise only. 
We see that when feedback is disabled, the effective noise floor rises by about 3~dB. 
}
\end{figure}

This minimization of tone power on the cryogenic amplifier is critical 
to reducing intermodulation products in band, which may contribute to
crosstalk or noise. 
We demonstrate the improvement of the overall noise floor due to reduced 
intermodulation products in Fig~\ref{fig:tracking_onoff}. 
Here, a single NIST CMB-style \umux resonator chip with 65 SQUID-coupled 
resonators spanning 4.25-4.5~MHz is calibrated with SMuRF at a tone 
power roughly corresponding to -70~dBm per tone at the input of the resonators. 
After interacting with the resonators, the tones are amplified with a 
4K HEMT (Low Now Factory LNF-LNC4-8C-LG), a 50K follow-on amplifier 
(B\&Z Technologies BZUCR-04000800-031012-102323), and room-temperature 
amplifier (MiniCircuits ZX60-83LN12+) and input to a spectrum analyzer. 
After feedback is disabled, no additional external flux is applied to the SQUID. 
The measurement was limited by availability of cold multiplexer chips at the time 
it was performed; however, we see that even with a small subset of the full 
\Order(1000) tones used in typical operation, the overall noise floor due to 
the additional intermodulation products when feedback is disabled is elevated 
by $\sim~3$ dBm, thus degrading the RF dynamic range. 
This degradation is prohibitive for \umux at full multiplexing density; thus, the 
tone-tracking capability of SMuRF is key to achieving the \Order(1000)x or higher 
multiplexing factors desired. 

\subsection{\label{sec:perfbandwidth}Bandwidth}

\begin{figure}
\includegraphics[width=0.99\columnwidth]{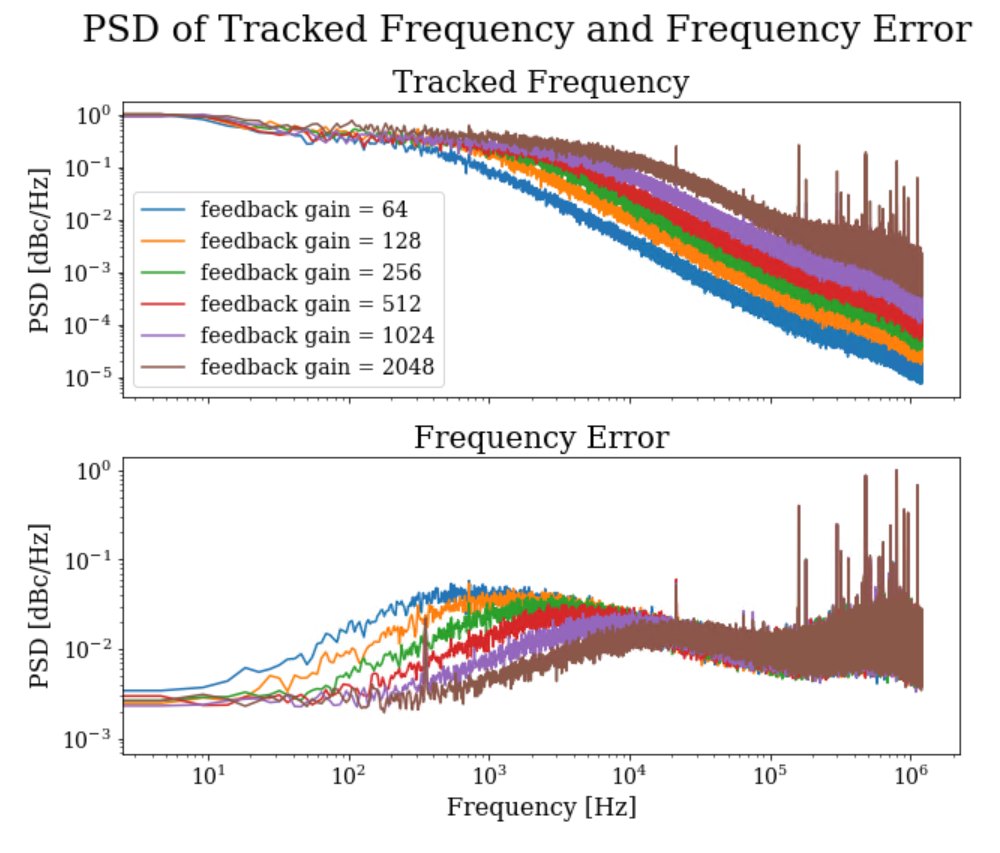}
\caption{\label{fig:trackedfreqspec} Power spectral density of tracked frequency 
and frequency error for varying gains and no flux ramp. We see that the tracked 
frequency rolls off at a few kHz or above depending on the gain setting. 
The frequency error is suppressed at low frequencies, corresponding to the majority 
of the power being captured by the tracked frequency.}
\end{figure}

The measurements in \S~\ref{sec:perfdyn} and \ref{sec:perftracking}
indicate that the SMuRF achieves the synthesis and analysis of tones
across the full 4-8~GHz bandwidth while exceeding the requirement for 
RF dynamic range. 
We focus now on the achievable bandwidth of the detector readout
firmware, including tone tracking and flux ramp demodulation.
A discussion of the full bandwidth performance of the tone-tracking
algorithm is deferred to a future publication.
However, we see in Figure~\ref{fig:trackedfreqspec} that in the ``DC
tracking'' state in which the flux ramp has not been activated, the
power spectral density of tracked frequency and frequency error rolls
off between $1$ and $10$~kHz depending on the gain settings used.

\begin{figure}
\includegraphics[width=0.99\columnwidth]{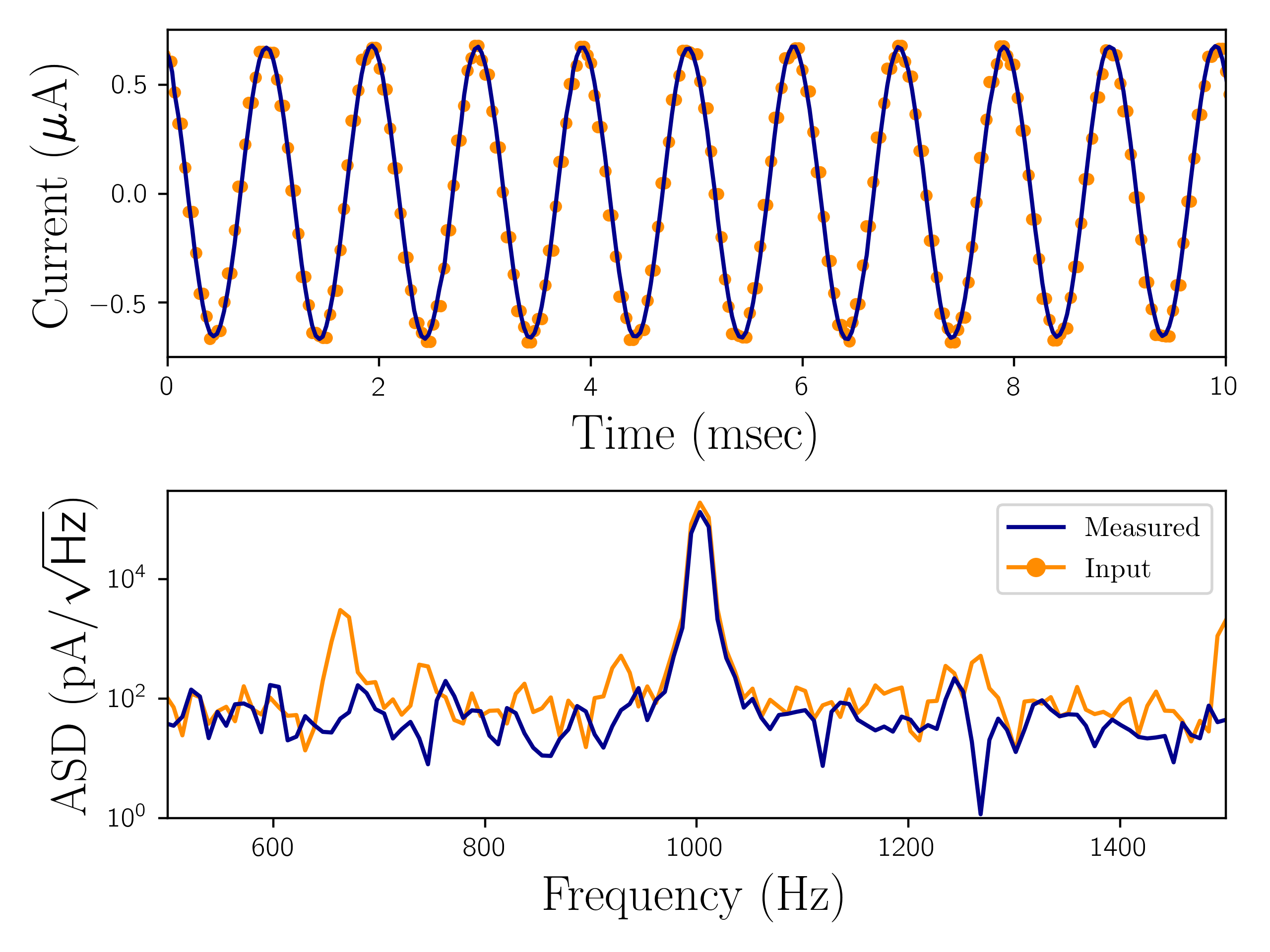}
\caption{\label{fig:trackedsine} [Top] Input signal and flux ramp
  demodulated signal played on the detector bias line while the
  detector was superconducting. 
  The raw input signal is sparsely sampled and scaled and shifted to account 
  for the transfer function of the superconducting detector's bias circuit 
  and latency.  
  [Bottom] Amplitude spectral density (ASD) of the input and measured
  signals. 
  SMuRF easily generates and recovers a 1~kHz sine wave, here
  sampled at a flux ramp frame rate of 30~kHz with 1~$\Phi_0$ per
  ramp. 
  The white noise floor of $\sim100$~pA/$\sqrt{\mathrm{Hz}}$ is due to 
  Johnson noise generated by the superconducting detector's cold bias resistor.
}
\end{figure}

Using the adaptive feedback algorithm with tone tracking, previous measurements 
have indicated that the SMuRF is capable of tracking a flux ramp modulated 
resonance to flux ramp rates approaching 100~kHz on NIST CMB-style \umux resonators, 
well in excess of requirements for CMB polarimetry measurements.\cite{ltd21} 
We show explicitly in Figure~\ref{fig:trackedsine} that a 1~kHz sine wave injected 
on the detector bias line is resolved through the full tone tracking and flux ramp 
demodulation scheme. 
This measurement was limited by the ability of RTM slow DACs to play a pre-programmed 
waveform on the TES bias line, but is not near the capacity of the system. 

\subsection{\label{sec:perfcross}Crosstalk}
Since the polyphase filter bank achieves 100~dBc/Hz SFDR, it is expected that 
the crosstalk between channels is dominated by the hardware components. 
The linearity measurements from \S~\ref{sec:perftracking} indicate that the 
impact of additional probe tones being turned on in other channels does not 
contribute meaningfully to the measured characteristics of existing channels. 

Previous measurements with SMuRF in a full field deployment setting including 
telescope scanning and streaming to external DAQ indicate that the crosstalk 
between non-pathological cold multiplexer channels integrated with TESs is at 
least 30~dB below the primary detector channel.\cite{yu19} 

More recent laboratory measurements of the cold multiplexer only constrain the 
crosstalk of well-behaved channels to about 1 part in $10^4$, consistent with 
expectations for the crosstalk of \umux cold hardware.\cite{dober21} 
These suggest that the crosstalk of the electronics comfortably meets the 
existing requirements and dominated by the cryogenic elements. 

\subsection{\label{sec:perfaltitude}Altitude}
We demonstrate that the SMuRF system is operable in high-altitude environments 
despite its computationally intensive digital signal processing algorithm. 

Due to high FPGA use and the 614.4~MHz maximum FPGA clock frequency, each SMuRF 
system (not including peripherals) consumes roughly 210 Watts.  
Out of this 210~W total, a typical SMuRF system draws 110~W for the SMuRF carrier, 
40~W for each AMC, and 20~W for the RTM.  
The power consumed by the SMuRF carriers is dominated by the main Xilinx Kintex 
UltraScale+ XCKU15P FPGA, which can draw as much as 80~W in steady state when 
reading out resonators over the full 4~GHz SMuRF bandwidth.  
The majority of the power drawn by the FPGA is consumed by the FPGA’s 0.85V 
internal supplies, which draw more than 50A which is provided by an 80A Intel 
Enpirion Power Solutions EM2280P01QI regulator on the SMuRF carrier.  
At these large current and power draws, the temperature limits of this regulator 
and the main FPGA are critical design considerations, particularly because the 
SMuRF systems are air cooled and intended for use at high altitude astronomical 
sites where the air pressure can be as much as 50\% lower than sea level. 
For the regulator, this is compounded by the fact that the regulator efficiency 
degrades near its 80A limit, which can result in thermal runaway if not carefully 
controlled for in design.  
FPGA lifetime is an additional concern, since expected lifetime can degrade if the 
FPGA temperature exceeds the Xilinx temperature specification.  
The maximum regulator operating temperature is 120\degr C.\footnote{\url{
https://www.xilinx.com/products/silicon-devices/fpga/kintex-ultrascale-plus.html\#documentation}}   
Xilinx specifies that the FPGAs must not exceed a temperature of 100\degr C 
for a 50A input current to ensure a >~10 year lifetime. 

Early SMuRF prototypes exceeded these temperature limits in simulated high 
altitude conditions.
This was addressed in the subsequent production systems by modifying the SMuRF 
FPGA firmware to significantly decrease resource usage and by making significant 
improvements to the FPGA and regulator heatsink designs and heatsink gluing 
materials and procedures.  
To verify performance, six full SMuRF systems, each consisting of a SMuRF carrier 
with two installed AMCs and an RTM, were tested in a 7-slot COMTEL ATCA crate 
along with all required peripherals, including a Dell R440 server and Xilinx 
KCU1500 data acquisition PCIe card, in a low pressure chamber.\footnote{Westpak, 
Inc. San Jose, CA, \url{https://www.westpak.com/}}
With the chamber pressure pumped down to the equivalent air pressure at 17,000~ft 
(395 Torr) to simulate operating conditions at a high altitude astronomical site, 
a test routine was run simulating the full computational load expected for the 
read out of the maximum 3328 channels over the full 4 GHz SMuRF bandwidth for each 
of the 6 SMuRF systems simultaneously.  
The chamber temperature was ramped from -5\degr C to +25\degr C and then back down to 
-5\degr C on a 3 hour cadence 120 times over the course of 14 days, simulating the 
typical diurnal temperature swing expected at a typical high altitude astronomical 
site based on environmental data from the high Andes.  
No subsystem (including peripherals) failures were observed over this 14 day run, 
and the SMuRF subsystem temperatures and currents were logged while ramping the 
chamber temperature.  
At 20\degr C ambient temperature, the averaged logged temperature for the SMuRF carrier 
FPGAs and regulators during this test was 71\degr C and 67\degr C, respectively, and 
the average regulator current at 20\degr C was 57A. 
The measured regulator temperatures are well below their 120\degr C operating limit, 
and the FPGA currents and temperatures meet Xilinx’s 10 year lifetime specification 
modified to account for these lower FPGA temperatures.\footnote{Xilinx, private 
communication}
The system was also successfully cold booted at -5\degr C three times, simulating 
remote recovery in winter conditions.

\subsection{\label{sec:perfacou}Vibration Sensitivity}

The acoustic sensitivity of the SMuRF is determined primarily by the
specific details of a given installation, such as the RF cabling used
and the mounting between the crate and cryostat. 
While a comprehensive frequency sweep has not been exhaustively probed, 
prior field deployment on a scanning telescope indicated the SMuRF
electronics suffered from no obvious additional noise due to the
telescope motion, indicating that mounting schemes exist such that the
SMuRF is sufficiently acoustically isolated.\cite{cukierman19}
Numerous existing lab measurement setups indicate no detectable 
acoustic sensitivity in the electronics from normal cryogenic laboratory 
settings, including the nearby operation of a pulse tube compressor 
and water chiller. 

In order to sufficiently cool the boards within a crate, which may
house arbitrarily many (typically up to 6) parallel SMuRF systems, we
run the SMuRF crate fans at high speeds. 
This fan motion does not appear to cause additional noise in the electronics.

\section{\label{sec:con}Conclusion and Future Outlook}

We have developed a digital readout and control system for use with
large-scale microwave resonator-based arrays.
The SMuRF electronics achieve > 100~dBc/Hz dynamic range at 30~kHz offset for 2000
simultaneous channels across its full 4-8~GHz bandwidth, with capacity
to read out up to 3328 channels across its readout bandwidth.
Each channel may return data with bandwidth up to 2.4~MHz and with a 
variety of processing options up to reporting flux ramp demodulated timestreams. 
Compared to other microwave resonator readout systems, SMuRF is unique
in its implementation of a tone-tracking algorithm, which allows for
dramatically increasing the channel count without exceeding the
linearity requirements of RF elements such as cryogenic amplifiers.
The system has been designed for integration with other data acquisition and 
experimental control platforms, and has been successfully integrated with 
control suites for the BICEP/Keck and Simons Observatory CMB experiments. 
It has been fielded previously within the BICEP/Keck program and is baselined 
for use on the upcoming Simons Observatory and BICEP Array experiments. 

Current efforts are underway to reconfigure the existing SMuRF
firmware and hardware architectures for KID systems by removing 
hardware and firmware functionality required only for \umux readout.
It is estimated that for resonators fabricated with sufficient
frequency accuracy, a single SMuRF card has the resources to support
20,000 KID-type resonators.
This would allow for megapixel arrays to be read out with a single
rack of equipment.

Future SMuRF development efforts include development of on-board pulse
processing algorithms for high-bandwidth applications, improving the
resonator tracking calibration scheme for asymmetric resonances, and
porting the existing software and firmware architecture to newly
available ``RF system on a chip'' (RFSoC) systems.
RFSoC systems allow for a large simplification in RF hardware design,
compact profile, and lower power consumption, making them particularly
attractive for balloon- and space-based missions.
Preliminary results on a Xilinx ZCU111 evaluation board demonstrated
comparable noise performance on 237 tracked channels, with the measurement
limited by the availability of cold resonators at the
time.\cite{dewart19}
In addition to reproducing the existing SMuRF firmware and performance
on the RFSoC, a number of improvements using the RFSoC platform are
being pursued.
We have developed new approaches to calibrating and improving the 
linearity of baseband I/Q modulation and demodulation schemes that provide 
the performance required for \umux without requiring the large band-splitting 
cavity filters present on the current AMCs. 
This technique has been successfully demonstrated with test code on RFSoC 
hardware and currently is being fully built out and tested.

\begin{acknowledgments}
CY was supported in part by the National Science Foundation Graduate Research 
Fellowship Program under Grant No. 1656518. 
MSF was supported in part by the Department of Energy (DOE) Office of Science Graduate 
Student Research (SCGSR) Program. 
The SCGSR program is administered by the Oak Ridge Institute for Science and 
Education (ORISE) for the DOE, which is managed by ORAU under contract number 
DE-SC0014664. 
ZA was supported in part by the DOE Office of Science Early Career Research Program. 
Several figures in this paper were generated with \texttt{numpy} and 
\texttt{matplotlib}\cite{numpy,hunter07}.
This work was supported by the Department of Energy at SLAC National Accelerator 
Laboratory under contract DE-AC02-76SF00515. 
\end{acknowledgments}

\section*{\label{sec:dataavailability}Data Availability Statement}
The data that support the findings of this study are available from the corresponding author upon reasonable request. 

\bibliography{smurf}

\end{document}